\documentclass[a4paper, amsfonts, amssymb, amsmath, reprint, showkeys, nofootinbib, twoside,superscriptaddress]{revtex4-2}
\usepackage[utf8]{inputenc}
\usepackage{graphicx} 
\usepackage{braket}
\usepackage{siunitx}
\usepackage{amsmath}
\usepackage{amsthm}
\usepackage{physics}

\begin{document}
\title{Gate-defined Two-dimensional Hole and Electron Systems in an Undoped InSb Quantum Well}

\author{Zijin Lei}
\email{zilei@phys.ethz.ch}
\affiliation{Solid State Physics Laboratory, ETH Zurich, CH-8093 Zurich, Switzerland}

\author{Erik Cheah}
\affiliation{Solid State Physics Laboratory, ETH Zurich, CH-8093 Zurich, Switzerland}

\author{Filip Krizek}
\affiliation{Solid State Physics Laboratory, ETH Zurich, CH-8093 Zurich, Switzerland}
\affiliation{IBM Research Europe - Zurich, Säumerstrasse 4, 8803 Rüschlikon, Switzerland}

\author{Rüdiger Schott}
\affiliation{Solid State Physics Laboratory, ETH Zurich, CH-8093 Zurich, Switzerland}

\author{Thomas Bähler}
\affiliation{Solid State Physics Laboratory, ETH Zurich, CH-8093 Zurich, Switzerland}

\author{Peter Märki}
\affiliation{Solid State Physics Laboratory, ETH Zurich, CH-8093 Zurich, Switzerland}

\author{Werner Wegscheider}
\affiliation{Solid State Physics Laboratory, ETH Zurich, CH-8093 Zurich, Switzerland}

\author{Mansour Shayegan}
\affiliation{Department of Electrical and Computer Engineering, Princeton University, Princeton, New Jersey 08544, USA}

\author{Thomas Ihn}
\affiliation{Solid State Physics Laboratory, ETH Zurich, CH-8093 Zurich, Switzerland}
\affiliation{ETH Zurich Quantum Center, CH-8093 Zurich, Switzerland}

\author{Klaus Ensslin}
\affiliation{Solid State Physics Laboratory, ETH Zurich, CH-8093 Zurich, Switzerland}
\affiliation{ETH Zurich Quantum Center, CH-8093 Zurich, Switzerland}

\begin{abstract}

Quantum transport measurements are performed in gate-defined, high-quality, two-dimensional hole and electron systems in an undoped InSb quantum well. For both polarities, the carrier systems show tunable spin-orbit interaction as extracted from weak anti-localization measurements. The effective mass of InSb holes strongly increases with carrier density as determined from the temperature dependence of Shubnikov–de Haas oscillations. Coincidence measurements in a tilted magnetic field are performed to estimate the spin susceptibility of the InSb two-dimensional hole system. The \textit{g}-factor of the two-dimensional hole system decreases rapidly with increasing carrier density. 
\end{abstract}

\maketitle

\section{Introduction}
Silicon, with both p-type and n-type transport characteristics, is the building block of today’s complementary metal–oxide–semiconductor (CMOS) circuits. However, not every semiconductor exhibits high quality for both carrier types, and it is even more challenging to achieve ambipolar operation in a single device. Nowadays, in the widely used silicon CMOS technology, the integration of p- and n-type components is often achieved with sophisticated fabrication techniques. For example, ion implantation is adopted to reverse the polarity on the wafer locally \cite{WILLIAMS19988,KUSTERS19919,mueller2015electron}. In contrast to the established processes for silicon, many challenges still remain in devices based on III-V compounds, whose mobilities are in general much higher. The p- and n-type components in III-V devices mostly rely on different doping methods and materials, which have to be combined with specific metals for realizing ohmic contacts \cite{stormer1983energy,sajoto1990fractional}. So far, most approaches rely on complex bilayer system designs, where p- and n-type layers are grown subsequently along the growth direction. There, the two-dimensional electron and hole gases (2DEG and 2DHG) are produced either by different choice of dopants, different semiconductor materials, or by electrostatic gating using both a top-gate and a back gate \cite{xiao2019anomalous,seamons2007undoped,kaestner2003quasi,nakajima2010high,karalic2017lateral,clarke2006fabrication,chu2016experimental}. Recently, incorporating both a high-quality 2DHG and a 2DEG into a single device became possible in 2D materials, for example, in graphene, MoS$_2$ or black phosphorous \cite{geim2009graphene,neto2009electronic,li2014black,chuang2014mos2}, since in these materials one can contact carriers in both conduction bands and valence bands \cite{zhan2014graphene,dean2012graphene,wang2019graphene,wang2012electronics,lee2014atomically}. 

Narrow-gap III-V compounds, such as InAs and InSb, have also been proposed as systems that can potentially host both p- and n-type polarities \cite{ashley2007heterogeneous,Doornbos2010}. InSb is known for the small effective mass of electrons and light holes in the bulk material, high carrier mobility at room temperature, strong spin-orbit interaction (SOI), and a large \textit{g}-factor in both the n- and p-type regimes \cite{vurgaftman2001band,kallaher2010spin,khodaparast2004spectroscopy,leontiadou2011experimental,gilbertson2009zero,orr2007surface,Madelung}. These unique properties are interesting for potential applications such as high-frequency electronics \cite{Ashley1995}, optoelectronics \cite{chen2005spin}, and spintronics \cite{vzutic2004spintronics}. More recently, InSb, together with InAs, has attracted attention as a potential platform for topological quantum information processing \cite{oreg2010helical,ke2019ballistic,deng2016majorana}. Various low-dimensional n-type InSb systems have been studied, including quantum wells \cite{Lehner2018,lei2019quantum}, nanowires \cite{Nilsson2009,Nilsson2010}, and other as-grown nanomaterials \cite{Mata2016twin,kang2018two,xue2019gate}. In previous work on n-type quantum wells we have shown that the parabolic band approximation agrees with experimental data \cite{Lei2020,Lei2021,Lei2022}. On the contrary, for lower-dimensional, p-type InSb, similar to other p-type III-V compounds, the band structures are predicted to be more complicated than their n-type counterparts. In general, they strongly depend on the dimensions, shape of the potential well, global or local strain fields, SOI, and the carrier confinement \cite{PEisenstein1984, Marcellina2017,Pryor2005}. Since the growth and processing methods for p-type devices are not as mature as for n-type, there is a lack of in-depth experimental studies of p-type InSb \cite{Gaspe2011}. So far, quantum transport measurements have been reported for p-type InSb nanowires \cite{Nilsson2011,Pribiag2013} and quantum wells \cite{Gaspe2011}. Also, room temperature experiments on p- and n-type InSb field-effect transistors were published \cite{radosavljevic2008high}. However, limited by both the material quality and device design, the important properties of p-type InSb, such as scattering mechanisms, effective mass, \textit{g}-factor, and SOI, have not been precisely characterized. 

In this work, we present quantum transport measurements on a gate-defined 2DHG and 2DEG in an ambipolar Hall bar device based on a single, undoped InSb quantum well. Both the 2DHG and the 2DEG have relatively high mobility, even though the quantum well is close to the surface to allow for electrostatic top-gating. Shubnikov–-de Haas (SdH) oscillations and the integer quantum Hall effect are observed for both carrier polarities. From the temperature dependence of SdH oscillations, we extract the effective mass of InSb holes and observe a large band non-parabolicity. Moreover, the strong and tunable SOI of both the 2DHG and the 2DEG is characterized by weak antilocalization (WAL) measurements. Finally, a coincidence measurement of the \textit{g}-factor in the 2DHG is performed, showing a strong dependence of the spin susceptibility on the carrier density. 

\section{Sample preparation}

\begin{figure}
    \centering
    \includegraphics[width=0.45\textwidth]{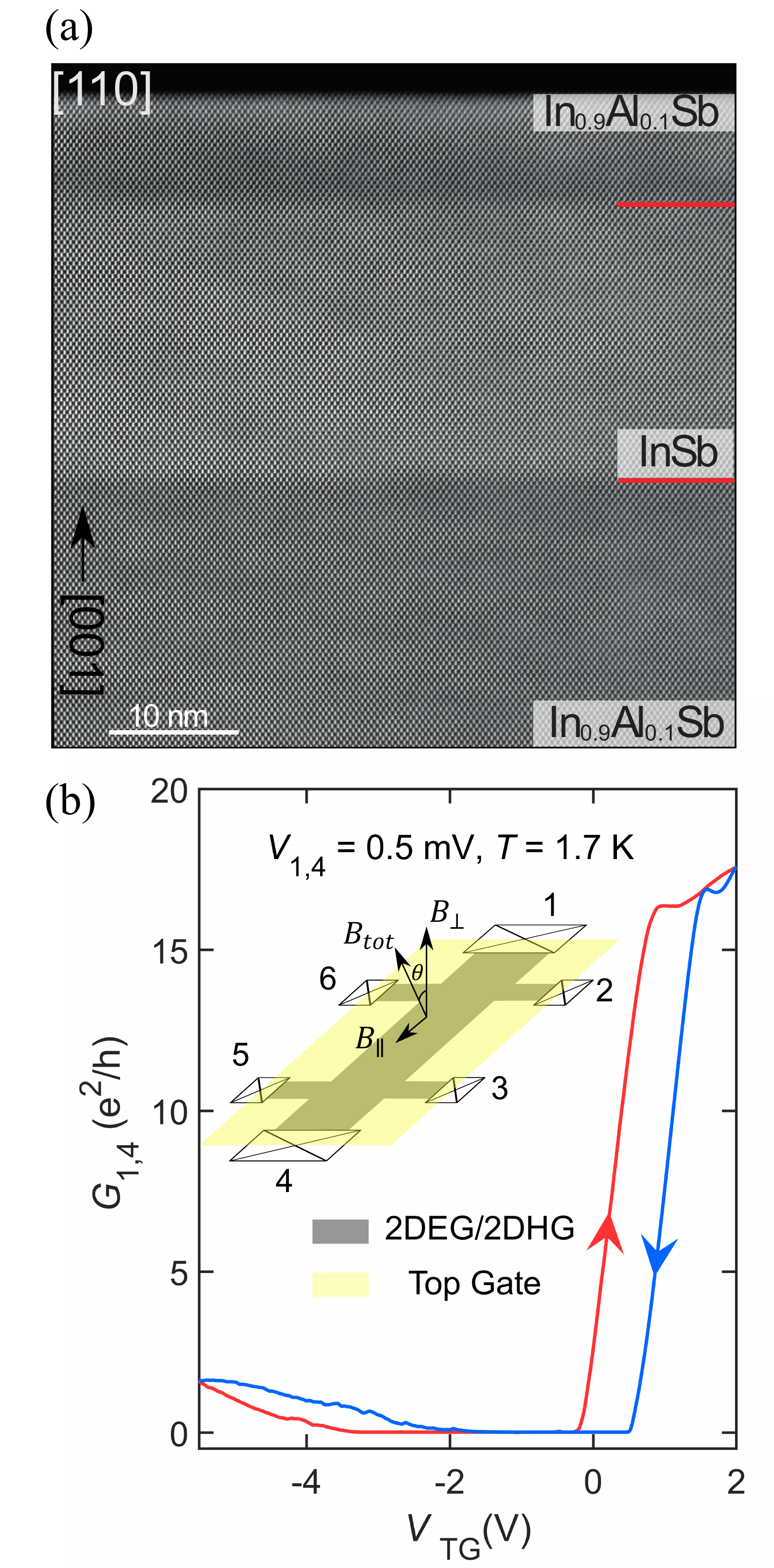}
    \caption{(a) High-angle-annular-dark-field scanning transmission electron microscopic (HAADF STEM) image of the InSb quantum well and neighboring In$_{0.9}$Al$_{0.1}$Sb barriers. There is no extended defects observed in the quantum well area. (b) Two-terminal conductance $G_{1,4}$ measured between contacts 1 and 4 vs. top-gate voltage $V_{\rm{TG}}$. A constant voltage of 0.5 mV is applied at 1.7 K. The blue and red traces show the down- and up-sweep, respectively. Inset: a schematic diagram of the Hall bar sample and the definition of tilt angle $\theta$. A global top-gate covers both the 2DEG or 2DHG and the contacts.}
    \label{fig1}
\end{figure}
The InSb quantum well sample we discuss here is grown on a GaAs (100) substrate by molecular beam epitaxy (MBE). The metamorphic buffer heterostructure, which is based on a previous publication by Lehner \textit{et al.} \cite{Lehner2018}, is used to filter the crystallographical defects related to the lattice mismatch. As shown in the high-angle-annular-dark-field (HAADF) scanning-transmission-electron-microscope (STEM) image of the
InSb quantum well sample in Fig. \ref{fig1} (a), a 21.5-nm-thick InSb quantum well is sandwiched in-between $\rm{In_{0.9}Al_{0.1}Sb}$ confinement barriers, with an 8.5-nm-thick top barrier and a 2-$\rm{\mu}$m-thick bottom barrier. We note that there is no intentional doping during growth. 

After defining a standard $400\times\SI{200}{\micro m^2}$ Hall bar using chemical etching, Ti/Au ohmic contacts are evaporated. This combination of metals has been reported to provide ohmic contacts for both p- and n-type InSb \cite{Nilsson2011,Pribiag2013}. In the next step, the sample is coated with a 40-nm-thick aluminum oxide dielectric layer using atomic layer deposition (ALD) at \SI{150}{\degree C}. Finally, a Ti/Au top-gate covering the Hall bar and overlapping the contacts is deposited by electron-beam evaporation, as illustrated in the inset of Fig. \ref{fig1}(b). 
\begin{figure}
    \centering
    \includegraphics[width=0.4\textwidth]{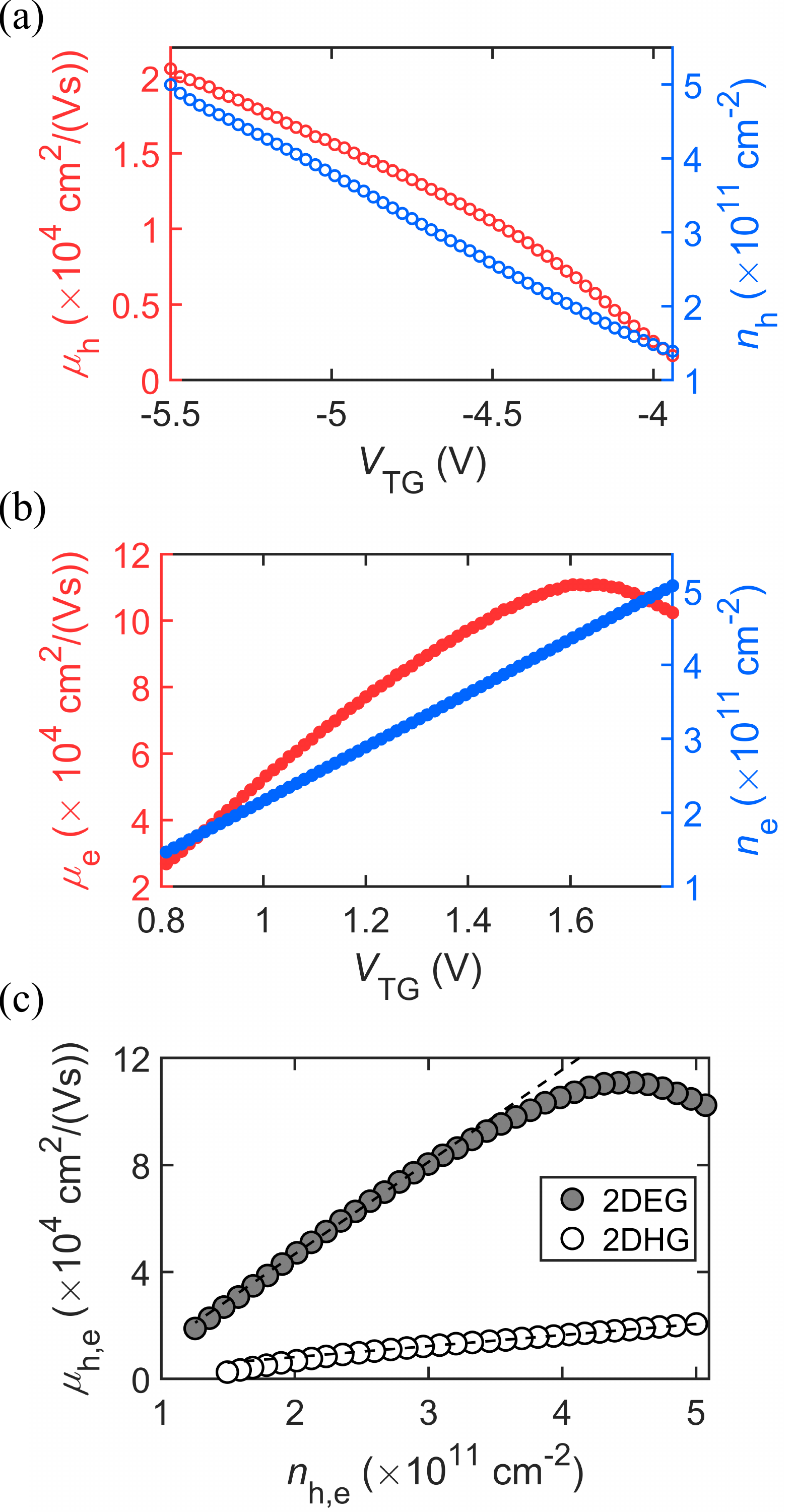}
    \caption{(a) and (b) Density and mobility vs. $V_{\rm{TG}}$ measured in p and n-type regimes, respectively. (c) The mobilities vs. densities for both p- and n-type. The dashed lines are linear fits. }
    \label{fig2}
\end{figure}

\begin{figure*}
    \centering
    \includegraphics[width=1\textwidth]{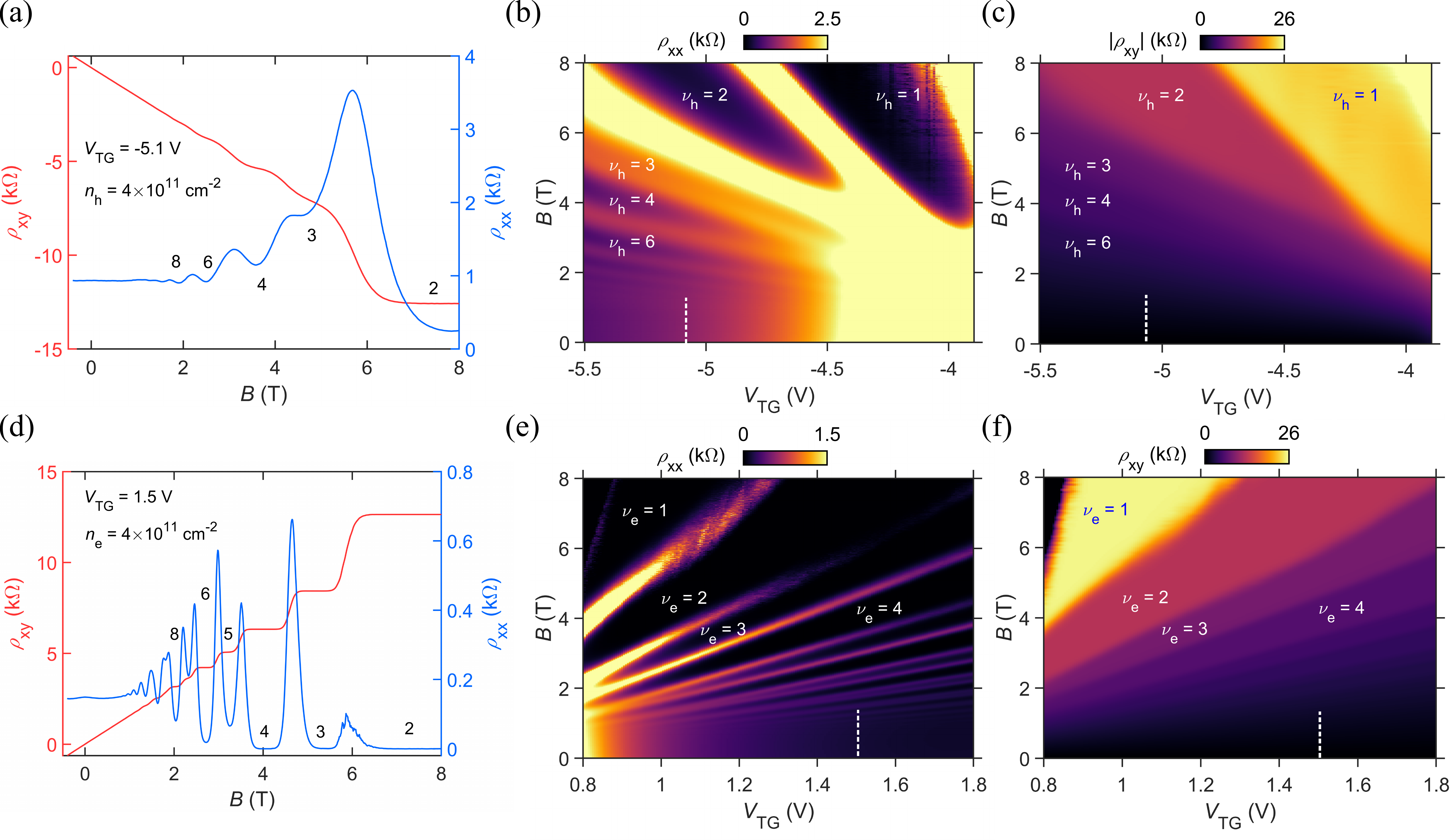}
    \caption{ (a) The SdH oscillations and integer quantum Hall effects of 2DHG. The filling factors are labeled in the figure. (b) and (c) The Landau fan diagrams of 2DHG. Both $\rho_{xx}$ and $\rho_{xy}$ are presented. (d), (e), and (f) Same as (a), (b), and (c), but in the n-type regime. The filling factors for 2DHG and 2DEG are labeled as $\nu_{\rm{h}}$ and $\nu_{\rm{e}}$, respectively. The white dashed lines correspond to the same carrier density $4\times10^{11}\: \rm{cm^{-2}}$, for both p- and n-types.}
    \label{fig3}
\end{figure*}

\section{SdH oscillations and integer quantum Hall effects}
Transport measurements are performed using standard low-frequency lock-in techniques in a cryostat with a base temperature of 1.7 K. Figure \ref{fig1}(b) presents the two-terminal conductance $G_{1,4}$ between contacts 1 and 4 as a function of top-gate voltage $V_{\rm{TG}}$. The conductance $G_{1,4}$ increases with increasing $V_{\rm{TG}}$ for positive voltages and with decreasing $V_{\rm{TG}}<\SI{-2}{V}$. In-between, the sample is insulating. This indicates that by sweeping $V_{\rm{TG}}$ from negative to positive, the polarity of the charge carriers changes from p-type to n-type, while in-between, the Fermi level traverses the bandgap. Comparing the up- and down-sweep of $V_{\rm{TG}}$ we find hysteresis. Therefore, we keep the sweep-direction fixed for the following measurements. In the n-type regime, down-sweeps of $V_{\rm{TG}}$ are used, while up-sweeps are used in the p-type regime. 

Now we measure the Hall bar sample with the standard method where a constant current is applied from contact 1 to contact 4. The longitudinal (between contact 2 and 3) and the transverse (between contact 3 and 5) resistivities, $\rho_{xx}$ and $\rho_{xy}$, are measured. Figures \ref{fig2}(a) and (b) show the densities $n_{\rm{h}}$ of holes and $n_{\rm{e}}$ of electrons, determined from the classical Hall resistance ($\rho_{xy}$) at small magnetic fields, and the corresponding Drude mobilities $\mu_{\rm{h}}$ and $\mu_{\rm{e}}$ obtained from $\rho_{xx}$ as a function of $V_{\rm{TG}}$. Figure \ref{fig2}(c) summarizes the data into plots of density vs. mobility for electrons and holes. The hole density $n_{\rm{h}}$ increases up to $\SI{5.0e11}{cm^{-2}}$ with decreasing $V_{\rm{TG}}$ and reaches the mobility $\mu_{\rm h}=\SI{21000}{cm^2/Vs}$. For comparison, also $n_{\rm{e}}$ increases up to $\SI{5e11}{cm^{-2}}$ with increasing $V_{\rm{TG}}$, but the corresponding $\mu_{\rm e}$ reaches its maximum of $\SI{1.1e5}{cm^2/Vs}$ at $n_{\rm{e}} = \SI{4.5e11}{cm^{-2}}$ and starts to decrease beyond. The decreasing electron mobility at increasing density suggests that interface roughness scattering starts to dominate.

Figures \ref{fig3}(a) and (d) present $\rho_{xx}$ and $\rho_{xy}$ vs. magnetic field $B$ in the hole and the electron regimes, where $V_{\rm{TG}}$ = $\SI{-5.1}{V}$ and $\SI{1.5}{V}$, respectively. The densities of both holes and electrons are $\SI{4e11}{cm^{-2}}$. We find pronounced SdH oscillations in $\rho_{xx}$ and quantum Hall plateaus close to $\rho_{xy}=h/\nu e^2$, where $\nu$ is the integer filling factor, in both the p-type and n-type regimes. The 2DEG and the 2DHG show a Zeeman splitting in the SdH oscillations starting at and below very different filling factors ($\nu=3$ for holes and $\nu=9$ for electrons), indicating significant differences in effective mass and \textit{g}-factors between electrons and holes. Related measurements are discussed further below. Landau fan diagrams of $\rho_{xx}$ and $\rho_{xy}$ for both 2DHG and 2DEG are presented in Figures \ref{fig3}(b,c) and (e,f), respectively. The corresponding filling factors for 2DHG and 2DEG are labeled as $\nu_{\rm{h}}$ and $\nu_{\rm{e}}$, respectively. Both regimes show the fan diagram of a single subband within the range of our measurements. Furthermore, for both 2DHG and 2DEG, there is no low-field, classical-positive magnetoresistance observed in $\rho_{xx}$ confirming that only a single subband is occupied for both polarities  \cite{lei2019quantum}. These measurements demonstrate the high quality of both the single-channel 2DEG and 2DHG electrically induced in the same device. Both of them are single-band systems, without the second electronic subband populated and without a noticeable splitting of the heavy-hole band due to SOI \cite{Winkler2003}. We can quantitatively estimate $\Delta n$, the carrier density at which the second electronic subband starts to be populated. Using the infinite-high quantum well model, $\Delta n$ is deduced to be $\sim\SI{10e11}{cm^{-2}}$, which is much larger than $n_{\rm{e}}$ and $n_{\rm{h}}$ in this work. 

\section{Effective mass and band non-parabolicity of 2D holes}
 \begin{figure}
    \centering
    \includegraphics[width=0.38\textwidth]{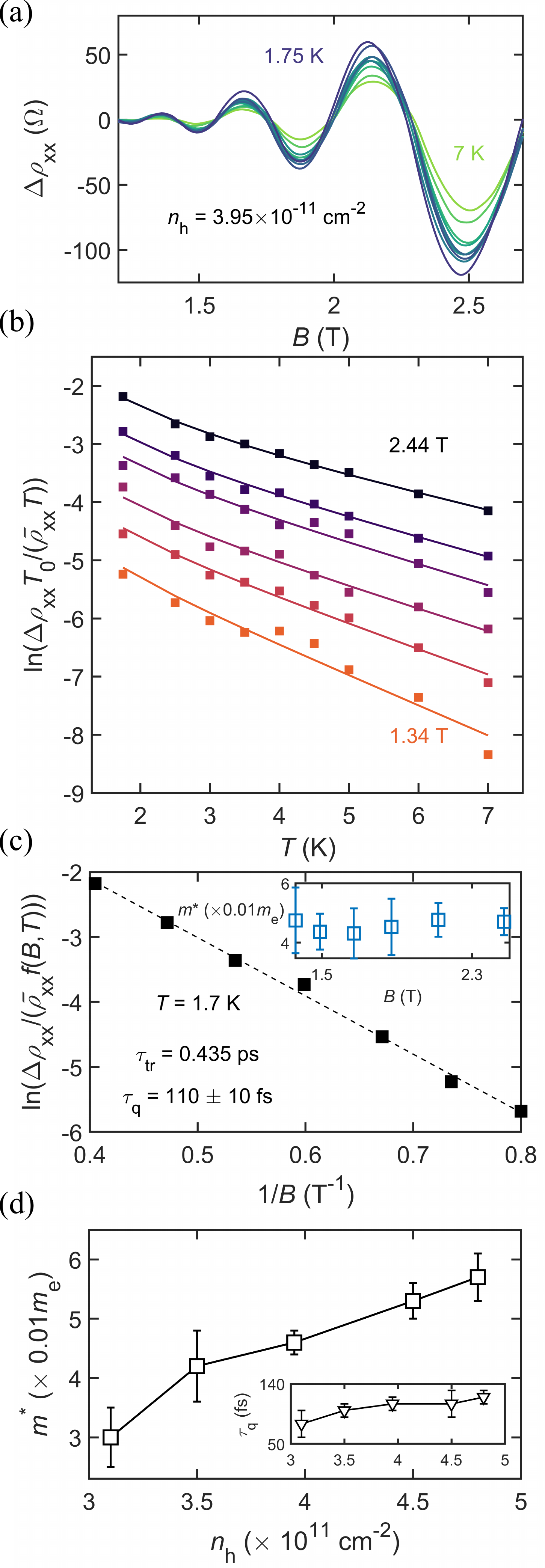}
    \caption{Effective mass measurement for 2DHG. (a) Temperature dependence of SdH oscillations with $n_{\rm{h}}= 3.95\times 10^{11}$ cm$^{-2}$. (b) Dingle factor fitting at different $B$. The squares are data and the lines are fitted curves. (c) The fitting of the quantum lifetime at 1.75 K. Inset: effective mass of holes vs. $B$ when $n_{\rm{h}}= 3.95\times 10^{11}$ cm$^{-2}$. (d) Effective mass as a function of carrier density. Error bars here are determined with the deviation of the fitting results. Inset: Dependence of $\tau_{\rm{q}}$ on $n_{\rm{h}}$ at 1.75 K.}
    \label{fig4}
\end{figure}
In our previous work, we have measured the effective mass of electrons in InSb quantum wells with the same thickness to be $\sim 0.016$ $m_{\rm{e}}$ \cite{lei2019quantum,Lei2020,Lei2022}. In the following, we present measurements of the effective mass of holes in the sample through the temperature dependence of the SdH oscillations at small magnetic fields. With the increase of the measurement temperature, the amplitude of the SdH oscillations will decrease until the oscillations vanish completely. The lower the effective mass, the higher the temperature at which the SdH oscillations can still be observed. The temperature dependence of the SdH oscillation is usually described by the Ando formula valid in the low magnetic field regime \cite{Ihn}.

Figure \ref{fig4}(a) shows SdH oscillations for a hole density $n_{\rm{h}} =\SI{3.95e11}{cm^{-2}}$, measured at temperatures from \SI{1.75}{K} to \SI{7}{K}. The oscillating part of the resistivity, $\Delta\rho_{xx}$, is obtained by subtracting the smooth background ($\bar{\rho}_{xx}$) from the measured $\rho_{xx}$. We select the local maxima and minima in the traces of $\Delta\rho_{xx}$ vs. $B$ to extract the temperature dependence of the SdH oscillations and plot the quantity $\ln{(\Delta\rho_{xx} T_0/ (\bar{\rho}_{xx} T))}$ as a function of temperature in Figure \ref{fig4}(b), where $T_0$ is the lowest temperature at which we measured the SdH oscillations. The effective mass $m^{\star}$ is then obtained by fitting to the function $-\ln(\sinh(2\pi^2m^\star k_\mathrm{B}T/\hbar eB))$ with the effective mass being the fitting parameter. The deduced effective mass of holes is $m^{\star}=0.046\pm 0.003$ $m_{\rm{e}}$, roughly constant within the magnetic field range of the measurement (inset of Fig. \ref{fig4}(c)).

The quantum lifetime was extracted by fitting the data in Fig. \ref{fig4}(c) with the Dingle factor and using the effective mass determined above. At \SI{1.75}{K}, we get a quantum lifetime of $\tau_{\rm{q}} = 110\pm10\:\rm{fs}$ from the slope of the linear fit $\ln{\Delta \rho_{xx}/(\Delta\bar{\rho}_{xx} ) f(B,T)}$ vs. $1/B$, where $f(B,T)=(2\pi^2 k_B T/\hbar\omega_c)/\sinh{(2\pi^2 k_B T/\hbar \omega_c)}$.
The quantum lifetime $\tau_{\rm{q}}$ is roughly a constant within the temperature range investigated. Considering the scattering time $\tau_{\rm{tr}} = 0.435$ ps deduced from the Drude model at this carrier density, the Dingle ratio is $\tau_{\rm{tr}}/\tau_{\rm{q}}\approx 4$. With the same carrier density in the n-type regime, the Dingle ratio is about 20.8 in this device. The large difference of the Dingle ratio may imply a large difference in the scattering mechanism in 2DHG and 2DEG. 

We repeat the measurement of the effective mass at different hole densities. In Fig. \ref{fig4}(d), $m^{\star}$ is plotted as a function of $n_{\rm{h}}$. With the increase of $n_{\rm{h}}$ from 3.1 to $\SI{4.85e11}{ cm^{-2}}$, the Fermi level shifts deeper into the valence band. At the same time, $m^{\star}$ increases from 0.03 to 0.057 $m_{\rm{e}}$. This indicates a very large band non-parabolicity. Compared to the effective mass of holes in InSb bulk material, our results show a slightly larger value than for the light holes (0.015 $m_{\rm{e}}$) but a much smaller value than the effective mass of heavy holes ($\sim0.43$ $m_{\rm{e}}$). This indicates that the band structure of p-type InSb may have similarities to p-type GaAs with the light-hole heave-hole degeneracy at the $\Gamma$ point being lifted due to the confinement in the growth direction. Theory predicts that the in-plane effective mass of the heavy-hole band is smaller than that of the light-hole band (mass inversion) \cite{Winkler2003}. The effective mass depends strongly on density, which either implies that bands are strongly altered by self-consistent effects, or strongly non-parabolic bands, or both \cite{Marcellina2017}. Compared with previous publications, our $m^{\star}$ are within the range of the results obtained through theoretical calculations and cyclotron resonance measurements \cite{Pryor2005,Gaspe2011}. Nevertheless, the value of $m^{\star}$ for holes depends on many factors, including the shape and width of the quantum wells, the strain, the carrier densities, and possibly the SOI.
Using the measured $m^{\star}$, we also obtain $\tau_{\rm{q}}$ as a function of $n_{\rm{h}}$ (inset of Fig. \ref{fig4}(d)), showing an increase of $\tau_{\rm{q}}$ with increasing carrier density, similar to mobility [Fig. \ref{fig2}(c)].

Using the values determined for $m^{\star}$, we qualitatively estimate the background impurity density $n_{\rm{imp}}$ in the heterostructure. Since $\mu_{\rm{h(e)}}$ is proportional to $n_{\rm{h(e)}}$ in a wide range, we may assume that scattering by ionized background impurities is the dominant scattering mechanism for both p- and n-type conduction \cite{umansky1997extremely}. 
Therefore, we estimate $n_{\rm{imp}}$ with 
\begin{equation}
\begin{split}
    \frac{1}{\tau_{\rm{tr}}} & \approx \frac{n_{\rm{imp}} m^{\star}}{2\pi\hbar^3 k_{\rm{F}}^3 } (\frac{e^2}{2\epsilon_0 \epsilon_{\rm{InSb}} })^2 \\
    & \int_{0}^{2k_F} \frac{qdq}{(q+q_{\rm{TF}})^2 \sqrt{1-(q/2k_{\rm{F}})^2}},
\end{split}
   \label{nimp}
\end{equation}
where $q_{\rm{TF}}$ is the Thomas--Fermi wave vector, $\epsilon_{\rm{InSb}}$ is the dielectric constant of InSb, and $\tau_{\rm{tr}}$ is the mean scattering time due to background impurity scattering \cite{Davies1997,Chung2022}. We calculate $n_{\rm{imp}}$ in the range where mobilities are proportional to carrier densities for 2DHG and 2DEG, respectively. $n_{\rm{imp}}$ is between $\SI{1.7e17}{cm^{-3}}$ and $\SI{3.8e17}{cm^{-3}}$ when $n_{h}$ is in a range of $3\sim\SI{4.7e11}{cm^{-2}}$. And when $n_{h}$ is in a range of $1.2\sim\SI{3e11}{cm^{-2}}$, $n_{\rm{imp}}$ is between $\SI{1.3e17}{cm^{-3}}$ and $\SI{1.8e17}{cm^{-3}}$. This value is slightly higher than previous publications on InSb quantum wells \cite{yi2015gate,Lei2022} and InAs quantum wells \cite{Tschirky2017}. Then, it is worth comparing the mobilities of our 2DHG and 2DEG in the light of Eq.~\eqref{nimp} considering the different effective masses of electrons and holes at $n_{\rm{e}}=n_{\rm{h}}=\SI{3e11}{cm^{-2}}$. If $n_{\rm{imp}}$ is the same for both 2DHG and 2DEG, this equation predicts $\mu_{\mathrm{e}}/\mu_{\mathrm{h}} \approx 2.4$. However, the measured mobility ratio is $\sim 6.2$, showing that the mobility of 2DHG in our InSb quantum well is lower than this prediction. This is similar to the high-mobility GaAs 2D systems \cite{Chung2022,Ahn2022}, and more investigation is needed to understand the scattering mechanism in InSb 2D systems quantitatively. 
\section{SOI of 2DHG and 2DEG}
 \begin{figure}
    \centering
    \includegraphics[width=0.5\textwidth]{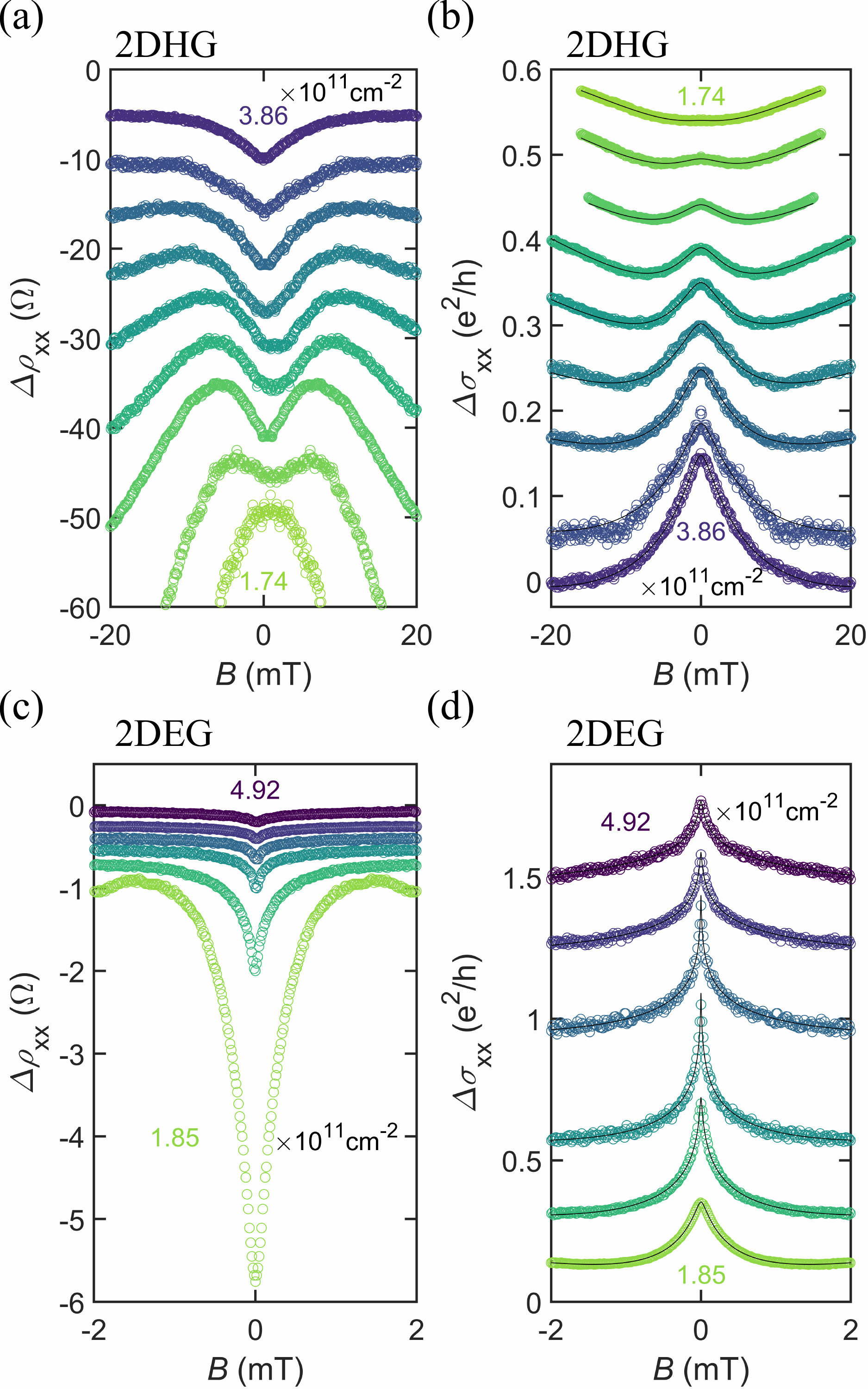}
    \caption{Density dependence of WAL for both 2DHG and 2DEG. (a) and (b) present $\Delta \rho_{xx}$ and $\Delta \sigma_{xx}$ vs. $B$ for 2DHG, respectively. Circles are the data while black curves are fits. (c) and (e). Same as (a) and (b), but for the 2DEG. There are vertical offsets added in all the figures here for a clear presentation. }
    \label{fig5}
    
\end{figure}

Both the high-quality 2DHG and 2DEG in undoped InSb quantum wells display tunable SOI, which we characterize by WAL measurements. Figure \ref{fig5} shows the WAL measurement at different carrier densities for the 2DHG and the 2DEG at \SI{1.7}{K}, respectively. For the convenience of comparison, we present $\Delta \rho_{xx}(B)=\rho_{xx}(B)-\rho_{xx}(0)$. To increase the signal-to-noise ratio, each trace presented represents the average of more than 5 consecutive measurements. In Fig. \ref{fig5}(a), with the increase of $n_{\rm{h}}$ from 1.74 to $\SI{3.86e11}{cm^{-2}}$, the local maximum of $\Delta \rho_{xx}$ at zero magnetic field develops into a minimum, which indicates the evolution from weak localization (WL) to WAL. Furthermore, in the WAL regime, $\Delta \rho_{xx}$ first increases and then decreases with the increase of the absolute value of $B$. At higher carrier densities, the turn-over magnetic field of $\Delta \rho_{xx}$ vs. $B$ moves to higher values. This implies an increase of the SOI with increasing $n_{\rm{h}}$. For the 2DEG, we find WAL throughout the whole range of carrier densities which is from 1.85 to $\SI{4.92e11}{cm^{-2}}$ (Fig. \ref{fig5}(c)). Here we point out that whether or not the 2D system is in the WL or WAL regimes mainly depends on the ratio between $l_{\rm{SO}}$ and the mean-free-path $l_{\rm{e}}$ \cite{Ihn}. In our device, $l_{\rm{SO}}/l_{\rm{e}}$ can be tuned over a larger range in the 2DHG than in the 2DEG. This is the reason why the evolution from the WL regime to the WAL regime is not observed in the 2DEG. 

\begin{figure}
    \centering
    \includegraphics[width=0.38\textwidth]{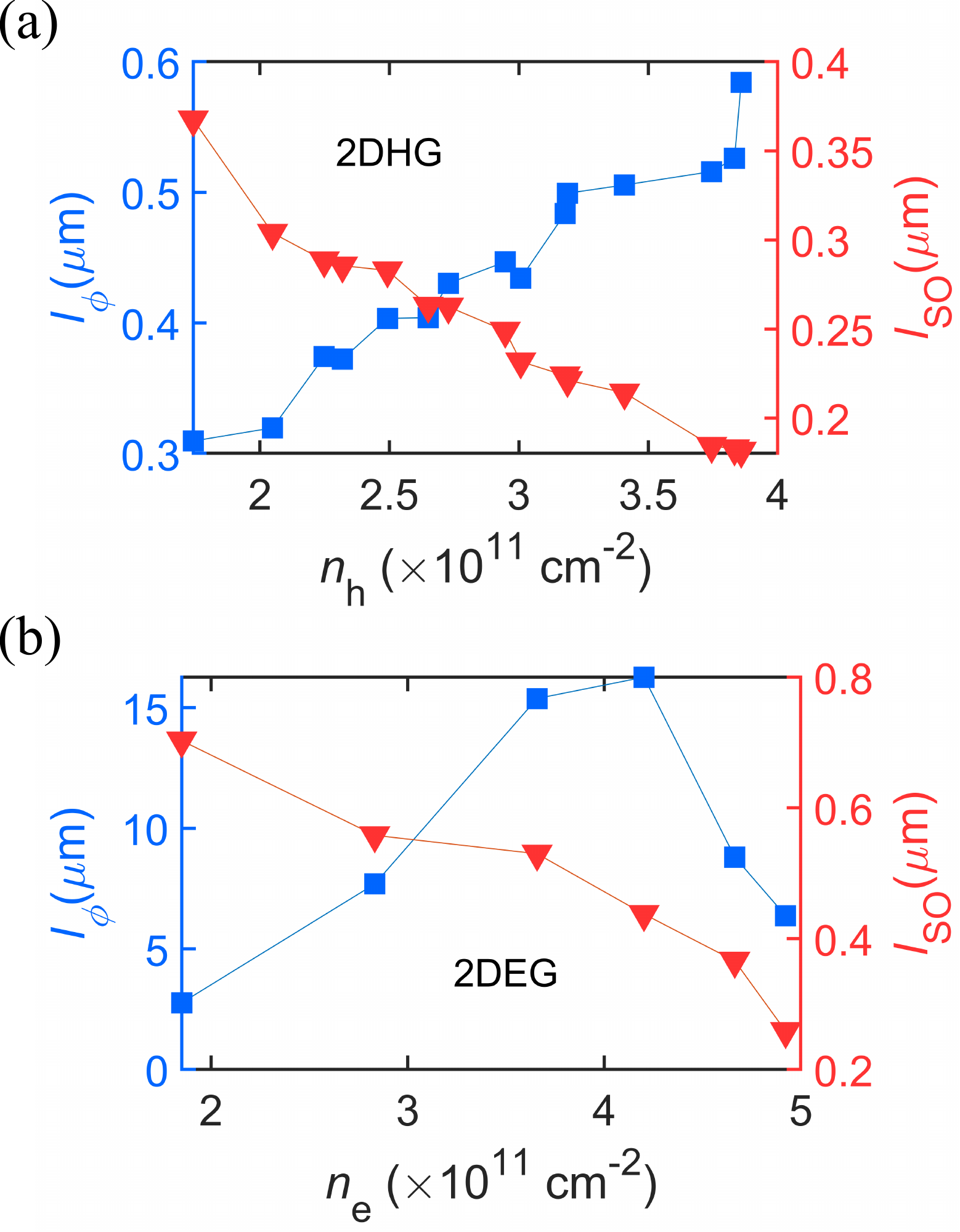}
    \caption{(a) $l_{\rm{SO}}$ and $l_\phi$ as a function of carrier density, are presented in (a) and (b) for 2DHG and 2DEG, respectively.}
    \label{fig6}
\end{figure}

To understand this process in-depth, we extract the coherence length and the spin-orbit length using the Hikami-Larkin-Nagaoka (HLN) expression \cite{Hikami1980,Iordanskii1994}. Similar methods were also adopted in publications where WAL effects were measured in doped InSb quantum wells \cite{ke2019ballistic}, InSb nanosheets \cite{chen2021strong}, and in our previous work \cite{Lei2022}. As shown in Figs. \ref{fig5}(b) and (d), we convert the $\rho_{xx}$ and $\rho_{xy}$ to longitudinal conductivity ($\sigma_{xx}$) and subtract a classical polynomial background from the measurement in both p- and n-type regimes. The conductivity correction $\sigma_{xx}$ of WALs reads:
\begin{equation}
    \begin{split}
        \Delta\sigma_{xx}(B)&=\frac{e^2}{2\pi^2\hbar}\left\{\Psi(\frac{1}{2}+\frac{H_\phi}{B}+\frac{H_{\rm{SO}}}{B})\right.\\ &+\frac{1}{2}\Psi(\frac{1}{2}+\frac{H_\phi}{B}+\frac{2H_{\rm{SO}}}{B})\\
        &-\frac{1}{2}\Psi(\frac{1}{2}+\frac{H_\phi}{B})-\ln{(\frac{H_\phi+H_{\rm{SO}}}{B})}\\
        &\left.-\frac{1}{2}\ln{(\frac{H_\phi+2H_{\rm{SO}}}{B})}+\frac{1}{2}\ln{\frac{H_\phi}{B}}\right\}.
    \end{split}
\end{equation}
Here, the two fitting parameters $H_\phi$ and $H_{\rm{SO}}$ are the phase-coherence field and the spin-orbit field, respectively, and $\Psi$ is the digamma function. The two fitting parameters can be converted to phase-coherence length $l_\phi$ and spin-orbit length $l_{\rm{SO}}$ using $l_\phi =\sqrt{\hbar/4eH_{\phi}}$ and $l_{\rm{SO}} =\sqrt{\hbar/4eH_{\rm{SO}}}$. We first focus on the 2DHG. As shown in Fig. \ref{fig6}(a), the phase coherence length $l_\phi$ increases from 0.31 to 0.59 $\rm{\mu}$m with the increase of $n_{\rm{h}}$. The reduction of decoherence agrees with the increase of mobility. In contrast to $l_\phi$, the spin-orbit length $l_{\rm{SO}}$ decreases from 0.37 to 0.18 $\rm{\mu}$m, which indicates an increase of the SOI with higher $n_{\rm{h}}$.

Like for other systems with strong SOI, we estimate the energy scales of the SOI with a parabolic band approximation \cite{Senz2000,Grbic2008}. Since we have determined the effective mass of the holes, $\Delta_{\rm{SO}}$, the energy splitting at the Fermi level due to SOI can be calculated according to $\Delta_{\rm{SO}}=\sqrt{2\hbar^2/\tau_{\rm{tr}}\tau_{\rm{SO}}}$. The spin-orbit time $\tau_{\rm{SO}}$ is calculated from $\tau_{\rm{SO}}=\hbar/(4eDH_{\rm{SO}})$. Here, $D=v_{\rm{F}}^2\tau_{\rm{tr}}/2$ is the diffusion constant in 2D with $v_{\rm{F}}$ being the Fermi velocity, which depends on $m^{\star}$. Then, we normalize $\Delta_{\rm{SO}}$ to get the spin--orbit coefficient $\alpha_{\rm{SO}}=\Delta_{\rm{SO}}/2k_{\rm{F}}$. The calculation shows that $\Delta_{\rm{SO}}$ increases from 1.37 to 1.74 meV, while the spin--orbit coefficient $\alpha_{\rm{SO}}$ is constant at $\sim \SI{53}{\milli\electronvolt\angstrom}$. The constant $\alpha_{\rm{SO}}$ is mainly due to the increasing mass. If we roughly estimate the Fermi energy using $E_{\rm{F}}=\hbar^2 k_{\rm{F}}^2/2m^{\star} $, the ratio $\Delta_{\rm{SO}}/E_{\rm{F}}$ increases from 5.2 $\%$ to 8.2 $\%$ in the range of $n_{\rm{h}}$ measured.

The results for the 2DEG are qualitatively similar to those for the 2DHG. Here, as shown in Fig. \ref{fig6}(b), $l_{\rm{SO}}$ decreases from 0.66 to 0.23 $\rm{\mu}$m, which implies an increasing SOI with increasing $n_{\rm{e}}$. For electrons, $\Delta_{\rm{SO}}$ increases from 0.73 to 3.2 meV, and $\alpha_{\rm{SO}}$ increases from 34 to $\SI{91}{\milli\electronvolt\angstrom}$. This value is approaching the results obtained recently in InAsSb quantum wells \cite{metti2022}. Interestingly, $l_\phi$ reaches a maximum of 16 $\rm{\mu}$m and decreases with higher $n_{\rm{e}}$. This is likely because of the mobility in the 2DEG, when the electron density $n_{\rm{e}}$ exceeds $\SI{4e11}{ cm^{-2}}$. 

\section{Coincidence measurements of 2DHG}

Finally, we present measurements of the spin susceptibility of the 2DHG using the coincidence method. This method has been introduced in previous publications \cite{Brossig2000,Lei2020}. The cyclotron energy is proportional to the perpendicular magnetic field $B_{\perp}$, while the Zeeman energy is proportional to the total magnetic field $B_{\rm tot}$, if an isotropic $g$-factor is assumed. The ratio between these two energies can be continuously changed by changing $\theta$, the angle between the directions of sample normal and total magnetic field $B_{\rm tot}$ [inset of Fig. \ref{fig1}(b)]. Therefore, $B_{\perp}$, the projection of the magnetic field onto the sample normal, is $B_{\perp}$ = $B_{\rm tot} \cos{\theta}$. Here, we introduce the parameter:
\begin{equation}
    r=\frac{g^{\star}\mu_{\rm B} B_{\rm tot}}{\hbar\omega_{\rm c}}=\frac{g^{\star}\mu_{\rm B} B_{\rm tot}}{e\hbar B_\perp/m^{\star}}
\end{equation}
to represent the ratio between Zeeman energy and cyclotron energy ($\mu_{\rm B}$ is the Bohr magneton). Hence, we obtain $r\cos{\theta}=g^{\star}m^{\star}/2m_{\rm{e}}$. Whenever $r$ takes integer values, SdH-minima in $\rho_{xx}$ turn into maxima because the energy gap between the Landau levels neighboring in energy vanishes. The value of $r$ can therefore be obtained by following the behavior of the local maxima and minima of $\rho_{xx}$ for changing tilt angles $\theta$. In general, for example in InAs and InSb 2DEGs, the value of $\rho_{xx}$ at integer filling factors will change periodically between local minima and maxima, with increasing $r$ \cite{SCHUMACHER1998,Brossig2000}. Knowing $r$ and $\theta$, one can estimate the spin susceptibility to be $\chi=(g^{\star} m^{\star})/(2\pi\hbar^2 )$. 

 \begin{figure}
    \centering
    \includegraphics[width=0.45\textwidth]{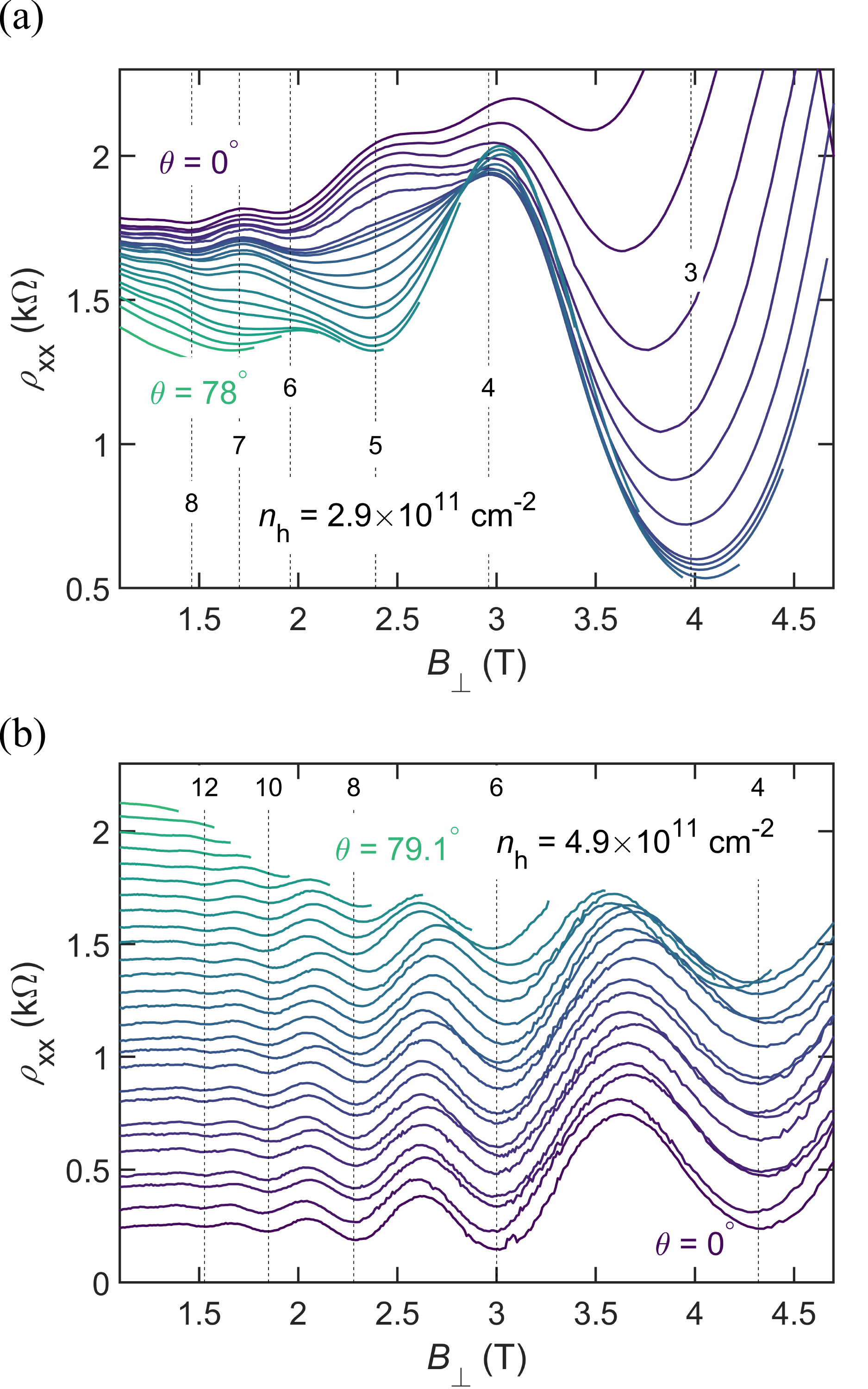}
    \caption{(a) and (b) The SdH oscillations measured for different tilt angles $\theta$ when $n_{\rm{h}} = 2.9$ and 4.9 $\times10^{11}$ cm$^{-2}$, respectively. For a better presentation, there is a constant vertical offset of 0.05 k$\Omega$ between each trace in (b) but no offset in (a). The filling factors are labeled with dashed lines.}
    \label{fig7}
\end{figure}

Figure \ref{fig7} shows $\rho_{xx}$ as a function of $B_{\perp}$ with $\theta$ varying from 0 to $\sim80^{\circ}$ for densities $n_{\rm{h}} = 2.9\mbox{ and }\SI{4.9e11}{ cm^{-2}}$. The tilt angle $\theta$ is calibrated precisely (within $0.1^{\circ}$) using the Hall effect at small $B_{\perp}$. As shown in Fig. \ref{fig7}(a), despite being limited by the range of the magnetic field, we observe local minima and maxima in $\rho_{xx}(B)$ to change as $\theta$ is increased. For instance, with increasing $\theta$, the local $\rho_{xx}$ maxima at $\nu_{\rm{h}} = 5$ and 7 change to minima, while the pronounced minima at $\nu_{\rm{h}}$ = 6 and 4 change to maxima or vanish. This means that $r$ is approaching 1 when we increase $\theta$ from 0 to $78^{\circ}$. As an estimation of the upper limit of the \textit{g}-factor, we assume $r\approx 1$ when $\theta\geq78^{\circ}$. This corresponds to a \textit{g}-factor $g^{\star} \leq 14$ for the measured $m^{\star}\approx0.03m_{\rm{e}}$. This estimation of the \textit{g}-factor is in agreement with the value in bulk p-type InSb, which is $\sim 16$ \cite{vurgaftman2001band,Winkler2003}. Here we also find that the local $\rho_{xx}$ minima for $\nu_{\rm{h}}=4$ and 3 appear in smaller $B_{\perp}$ when $\theta = 0$. This could be because of the effect of attractive scattering centers in the quantum well in the low density regime \cite{Raug1987}.  

However, coincidence measurements at higher densities show different results. Figure \ref{fig7}(b) presents the coincidence measurement performed at $n_{\rm{h}}=\SI{4.9e11}{cm^{-2}}$. Even when $\theta$ is increased to $79.1^{\circ}$, within the range of the magnetic field we can apply, there is no obvious dependence of $\rho_{xx}$ on $\theta$. Since p-type InSb is regarded to be a material with strong SOI and large \textit{g}-factor, it is unlikely that the \textit{g}-factor is so strongly reduced in the higher density regime that there is no observable change of $\rho_{xx}$ up to $\theta=79.1^{\circ}$. Since we have observed that the SOI of the 2DHG increases significantly with increasing $n_{\rm{h}}$, we hypothesize that the spin of the holes is locked with the orbits due to the larger SOI, implying a strongly anisotropic $g$-factor. Similar results have been reported in other p-type quantum wells and also in 2D materials, such as MoS$_2$ \cite{Shi2015,Pisoni2017}.

\section{Conclusion}
We have shown that both 2DHG and 2DEG can be induced by electrostatic gating in a single, shallow InSb heterostructure without any intentional doping. Thanks to the high quality of the sample, we are able to observe SdH oscillations and integer quantum Hall effects in both the p- and n-type regimes. In each regime, only a single electronic band is populated and contributes to transport. From the density dependence of the effective mass, we find that a large band non-parabolicity is present in the p-type regime. Furthermore, both 2DHG and 2DEG show tunable SOI. Finally, using coincidence measurements, we find that the spin susceptibility of InSb 2DHG strongly depends on the carrier density. Our work provides a better understanding of the detailed band structure and quantum properties of narrow-bandgap III-V compounds, and paves the way for applications of InSb in high-mobility electronics and quantum information processing devices.

\begin{acknowledgements}
We thank Marilyne Sousa and the Cleanroom Operations Team of the Binnig and Rohrer Nanotechnology Center (BRNC), IBM Research Zurich and Peng Zeng of the ScopeM, ETH Zurich for assistance during TEM characterization. We thank Shuichi Iwakiri, Lev Ginzburg, and Christoph Adam for fruitful discussions. We acknowledge funding from QuantERA. Mansour Shayegan acknowledges support through a QuantEmX travel grant from the Institute for Complex Adaptive Matter (ICAM) and the Gordon and Betty Moore Foundation through Grant No. GBMF5305. This work was supported by the Swiss National Science Foundation through the National Center of Competence in Research Quantum Science and Technology.
\end{acknowledgements}

\bibliography{bibliography.bib}

\begin{thebibliography}{72}%
\makeatletter
\providecommand \@ifxundefined [1]{%
 \@ifx{#1\undefined}
}%
\providecommand \@ifnum [1]{%
 \ifnum #1\expandafter \@firstoftwo
 \else \expandafter \@secondoftwo
 \fi
}%
\providecommand \@ifx [1]{%
 \ifx #1\expandafter \@firstoftwo
 \else \expandafter \@secondoftwo
 \fi
}%
\providecommand \natexlab [1]{#1}%
\providecommand \enquote  [1]{``#1''}%
\providecommand \bibnamefont  [1]{#1}%
\providecommand \bibfnamefont [1]{#1}%
\providecommand \citenamefont [1]{#1}%
\providecommand \href@noop [0]{\@secondoftwo}%
\providecommand \href [0]{\begingroup \@sanitize@url \@href}%
\providecommand \@href[1]{\@@startlink{#1}\@@href}%
\providecommand \@@href[1]{\endgroup#1\@@endlink}%
\providecommand \@sanitize@url [0]{\catcode `\\12\catcode `\$12\catcode
  `\&12\catcode `\#12\catcode `\^12\catcode `\_12\catcode `\%12\relax}%
\providecommand \@@startlink[1]{}%
\providecommand \@@endlink[0]{}%
\providecommand \url  [0]{\begingroup\@sanitize@url \@url }%
\providecommand \@url [1]{\endgroup\@href {#1}{\urlprefix }}%
\providecommand \urlprefix  [0]{URL }%
\providecommand \Eprint [0]{\href }%
\providecommand \doibase [0]{https://doi.org/}%
\providecommand \selectlanguage [0]{\@gobble}%
\providecommand \bibinfo  [0]{\@secondoftwo}%
\providecommand \bibfield  [0]{\@secondoftwo}%
\providecommand \translation [1]{[#1]}%
\providecommand \BibitemOpen [0]{}%
\providecommand \bibitemStop [0]{}%
\providecommand \bibitemNoStop [0]{.\EOS\space}%
\providecommand \EOS [0]{\spacefactor3000\relax}%
\providecommand \BibitemShut  [1]{\csname bibitem#1\endcsname}%
\let\auto@bib@innerbib\@empty
\bibitem [{\citenamefont {Williams}(1998)}]{WILLIAMS19988}%
  \BibitemOpen
  \bibfield  {author} {\bibinfo {author} {\bibfnamefont {J.}~\bibnamefont
  {Williams}},\ }\bibfield  {title} {\bibinfo {title} {Ion implantation of
  semiconductors},\ }\href
  {https://doi.org/https://doi.org/10.1016/S0921-5093(98)00705-9} {\bibfield
  {journal} {\bibinfo  {journal} {Materials Science and Engineering: A}\
  }\textbf {\bibinfo {volume} {253}},\ \bibinfo {pages} {8} (\bibinfo {year}
  {1998})}\BibitemShut {NoStop}%
\bibitem [{\citenamefont {Küsters}\ \emph {et~al.}(1991)\citenamefont
  {Küsters}, \citenamefont {Mühlhoff},\ and\ \citenamefont
  {Cerva}}]{KUSTERS19919}%
  \BibitemOpen
  \bibfield  {author} {\bibinfo {author} {\bibfnamefont {K.}~\bibnamefont
  {Küsters}}, \bibinfo {author} {\bibfnamefont {H.}~\bibnamefont
  {Mühlhoff}},\ and\ \bibinfo {author} {\bibfnamefont {H.}~\bibnamefont
  {Cerva}},\ }\bibfield  {title} {\bibinfo {title} {Application of ion
  implantation in submicron {CMOS} processes},\ }\href
  {https://doi.org/https://doi.org/10.1016/0168-583X(91)96126-6} {\bibfield
  {journal} {\bibinfo  {journal} {Nuclear Instruments and Methods in Physics
  Research Section B: Beam Interactions with Materials and Atoms}\ }\textbf
  {\bibinfo {volume} {55}},\ \bibinfo {pages} {9} (\bibinfo {year}
  {1991})}\BibitemShut {NoStop}%
\bibitem [{\citenamefont {Mueller}\ \emph {et~al.}(2015)\citenamefont
  {Mueller}, \citenamefont {Konstantaras}, \citenamefont {Spruijtenburg},
  \citenamefont {van~der Wiel},\ and\ \citenamefont
  {Zwanenburg}}]{mueller2015electron}%
  \BibitemOpen
  \bibfield  {author} {\bibinfo {author} {\bibfnamefont {F.}~\bibnamefont
  {Mueller}}, \bibinfo {author} {\bibfnamefont {G.}~\bibnamefont
  {Konstantaras}}, \bibinfo {author} {\bibfnamefont {P.~C.}\ \bibnamefont
  {Spruijtenburg}}, \bibinfo {author} {\bibfnamefont {W.~G.}\ \bibnamefont
  {van~der Wiel}},\ and\ \bibinfo {author} {\bibfnamefont {F.~A.}\ \bibnamefont
  {Zwanenburg}},\ }\bibfield  {title} {\bibinfo {title} {Electron--hole
  confinement symmetry in silicon quantum dots},\ }\href@noop {} {\bibfield
  {journal} {\bibinfo  {journal} {Nano letters}\ }\textbf {\bibinfo {volume}
  {15}},\ \bibinfo {pages} {5336} (\bibinfo {year} {2015})}\BibitemShut
  {NoStop}%
\bibitem [{\citenamefont {Stormer}\ \emph {et~al.}(1983)\citenamefont
  {Stormer}, \citenamefont {Schlesinger}, \citenamefont {Chang}, \citenamefont
  {Tsui}, \citenamefont {Gossard},\ and\ \citenamefont
  {Wiegmann}}]{stormer1983energy}%
  \BibitemOpen
  \bibfield  {author} {\bibinfo {author} {\bibfnamefont {H.}~\bibnamefont
  {Stormer}}, \bibinfo {author} {\bibfnamefont {Z.}~\bibnamefont
  {Schlesinger}}, \bibinfo {author} {\bibfnamefont {A.}~\bibnamefont {Chang}},
  \bibinfo {author} {\bibfnamefont {D.}~\bibnamefont {Tsui}}, \bibinfo {author}
  {\bibfnamefont {A.}~\bibnamefont {Gossard}},\ and\ \bibinfo {author}
  {\bibfnamefont {W.}~\bibnamefont {Wiegmann}},\ }\bibfield  {title} {\bibinfo
  {title} {Energy structure and quantized hall effect of two-dimensional
  holes},\ }\href@noop {} {\bibfield  {journal} {\bibinfo  {journal} {Physical
  review letters}\ }\textbf {\bibinfo {volume} {51}},\ \bibinfo {pages} {126}
  (\bibinfo {year} {1983})}\BibitemShut {NoStop}%
\bibitem [{\citenamefont {Sajoto}\ \emph {et~al.}(1990)\citenamefont {Sajoto},
  \citenamefont {Suen}, \citenamefont {Engel}, \citenamefont {Santos},\ and\
  \citenamefont {Shayegan}}]{sajoto1990fractional}%
  \BibitemOpen
  \bibfield  {author} {\bibinfo {author} {\bibfnamefont {T.}~\bibnamefont
  {Sajoto}}, \bibinfo {author} {\bibfnamefont {Y.}~\bibnamefont {Suen}},
  \bibinfo {author} {\bibfnamefont {L.}~\bibnamefont {Engel}}, \bibinfo
  {author} {\bibfnamefont {M.}~\bibnamefont {Santos}},\ and\ \bibinfo {author}
  {\bibfnamefont {M.}~\bibnamefont {Shayegan}},\ }\bibfield  {title} {\bibinfo
  {title} {Fractional quantum hall effect in very-low-density
  {G}a{A}s/{A}l$_x${G}a$_{1-x}${A}s heterostructures},\ }\href@noop {}
  {\bibfield  {journal} {\bibinfo  {journal} {Physical Review B}\ }\textbf
  {\bibinfo {volume} {41}},\ \bibinfo {pages} {8449} (\bibinfo {year}
  {1990})}\BibitemShut {NoStop}%
\bibitem [{\citenamefont {Xiao}\ \emph {et~al.}(2019)\citenamefont {Xiao},
  \citenamefont {Liu}, \citenamefont {Samarth},\ and\ \citenamefont
  {Hu}}]{xiao2019anomalous}%
  \BibitemOpen
  \bibfield  {author} {\bibinfo {author} {\bibfnamefont {D.}~\bibnamefont
  {Xiao}}, \bibinfo {author} {\bibfnamefont {C.-X.}\ \bibnamefont {Liu}},
  \bibinfo {author} {\bibfnamefont {N.}~\bibnamefont {Samarth}},\ and\ \bibinfo
  {author} {\bibfnamefont {L.-H.}\ \bibnamefont {Hu}},\ }\bibfield  {title}
  {\bibinfo {title} {Anomalous quantum oscillations of interacting
  electron-hole gases in inverted type-ii {I}n{A}s/{G}a{S}b quantum wells},\
  }\href@noop {} {\bibfield  {journal} {\bibinfo  {journal} {Physical review
  letters}\ }\textbf {\bibinfo {volume} {122}},\ \bibinfo {pages} {186802}
  (\bibinfo {year} {2019})}\BibitemShut {NoStop}%
\bibitem [{\citenamefont {Seamons}\ \emph {et~al.}(2007)\citenamefont
  {Seamons}, \citenamefont {Tibbetts}, \citenamefont {Reno},\ and\
  \citenamefont {Lilly}}]{seamons2007undoped}%
  \BibitemOpen
  \bibfield  {author} {\bibinfo {author} {\bibfnamefont {J.~A.}\ \bibnamefont
  {Seamons}}, \bibinfo {author} {\bibfnamefont {D.~R.}\ \bibnamefont
  {Tibbetts}}, \bibinfo {author} {\bibfnamefont {J.~L.}\ \bibnamefont {Reno}},\
  and\ \bibinfo {author} {\bibfnamefont {M.~P.}\ \bibnamefont {Lilly}},\
  }\bibfield  {title} {\bibinfo {title} {Undoped electron-hole bilayers in a ga
  as/ al ga as double quantum well},\ }\href@noop {} {\bibfield  {journal}
  {\bibinfo  {journal} {Applied Physics Letters}\ }\textbf {\bibinfo {volume}
  {90}},\ \bibinfo {pages} {052103} (\bibinfo {year} {2007})}\BibitemShut
  {NoStop}%
\bibitem [{\citenamefont {Kaestner}\ \emph {et~al.}(2003)\citenamefont
  {Kaestner}, \citenamefont {Wunderlich}, \citenamefont {Hasko},\ and\
  \citenamefont {Williams}}]{kaestner2003quasi}%
  \BibitemOpen
  \bibfield  {author} {\bibinfo {author} {\bibfnamefont {B.}~\bibnamefont
  {Kaestner}}, \bibinfo {author} {\bibfnamefont {J.}~\bibnamefont
  {Wunderlich}}, \bibinfo {author} {\bibfnamefont {D.~G.}\ \bibnamefont
  {Hasko}},\ and\ \bibinfo {author} {\bibfnamefont {D.}~\bibnamefont
  {Williams}},\ }\bibfield  {title} {\bibinfo {title} {Quasi-lateral
  2{DEG}--2{DHG} junction in {A}l{G}a{A}s/{G}a{A}s},\ }\href@noop {} {\bibfield
   {journal} {\bibinfo  {journal} {Microelectronics journal}\ }\textbf
  {\bibinfo {volume} {34}},\ \bibinfo {pages} {423} (\bibinfo {year}
  {2003})}\BibitemShut {NoStop}%
\bibitem [{\citenamefont {Nakajima}\ \emph {et~al.}(2010)\citenamefont
  {Nakajima}, \citenamefont {Sumida}, \citenamefont {Dhyani}, \citenamefont
  {Kawai},\ and\ \citenamefont {Narayanan}}]{nakajima2010high}%
  \BibitemOpen
  \bibfield  {author} {\bibinfo {author} {\bibfnamefont {A.}~\bibnamefont
  {Nakajima}}, \bibinfo {author} {\bibfnamefont {Y.}~\bibnamefont {Sumida}},
  \bibinfo {author} {\bibfnamefont {M.~H.}\ \bibnamefont {Dhyani}}, \bibinfo
  {author} {\bibfnamefont {H.}~\bibnamefont {Kawai}},\ and\ \bibinfo {author}
  {\bibfnamefont {E.~S.}\ \bibnamefont {Narayanan}},\ }\bibfield  {title}
  {\bibinfo {title} {High density two-dimensional hole gas induced by negative
  polarization at {G}a{N}/{A}l{G}a{N} heterointerface},\ }\href@noop {}
  {\bibfield  {journal} {\bibinfo  {journal} {Applied physics express}\
  }\textbf {\bibinfo {volume} {3}},\ \bibinfo {pages} {121004} (\bibinfo {year}
  {2010})}\BibitemShut {NoStop}%
\bibitem [{\citenamefont {Karalic}\ \emph {et~al.}(2017)\citenamefont
  {Karalic}, \citenamefont {Mittag}, \citenamefont {Tschirky}, \citenamefont
  {Wegscheider}, \citenamefont {Ensslin},\ and\ \citenamefont
  {Ihn}}]{karalic2017lateral}%
  \BibitemOpen
  \bibfield  {author} {\bibinfo {author} {\bibfnamefont {M.}~\bibnamefont
  {Karalic}}, \bibinfo {author} {\bibfnamefont {C.}~\bibnamefont {Mittag}},
  \bibinfo {author} {\bibfnamefont {T.}~\bibnamefont {Tschirky}}, \bibinfo
  {author} {\bibfnamefont {W.}~\bibnamefont {Wegscheider}}, \bibinfo {author}
  {\bibfnamefont {K.}~\bibnamefont {Ensslin}},\ and\ \bibinfo {author}
  {\bibfnamefont {T.}~\bibnamefont {Ihn}},\ }\bibfield  {title} {\bibinfo
  {title} {Lateral p- n junction in an inverted {I}n{A}s/{G}a{S}b double
  quantum well},\ }\href@noop {} {\bibfield  {journal} {\bibinfo  {journal}
  {Physical Review Letters}\ }\textbf {\bibinfo {volume} {118}},\ \bibinfo
  {pages} {206801} (\bibinfo {year} {2017})}\BibitemShut {NoStop}%
\bibitem [{\citenamefont {Clarke}\ \emph {et~al.}(2006)\citenamefont {Clarke},
  \citenamefont {Micolich}, \citenamefont {Hamilton}, \citenamefont {Simmons},
  \citenamefont {Muraki},\ and\ \citenamefont
  {Hirayama}}]{clarke2006fabrication}%
  \BibitemOpen
  \bibfield  {author} {\bibinfo {author} {\bibfnamefont {W.}~\bibnamefont
  {Clarke}}, \bibinfo {author} {\bibfnamefont {A.}~\bibnamefont {Micolich}},
  \bibinfo {author} {\bibfnamefont {A.}~\bibnamefont {Hamilton}}, \bibinfo
  {author} {\bibfnamefont {M.}~\bibnamefont {Simmons}}, \bibinfo {author}
  {\bibfnamefont {K.}~\bibnamefont {Muraki}},\ and\ \bibinfo {author}
  {\bibfnamefont {Y.}~\bibnamefont {Hirayama}},\ }\bibfield  {title} {\bibinfo
  {title} {Fabrication of induced two-dimensional hole systems on (311) a
  {G}a{A}s},\ }\href@noop {} {\bibfield  {journal} {\bibinfo  {journal}
  {Journal of applied physics}\ }\textbf {\bibinfo {volume} {99}},\ \bibinfo
  {pages} {023707} (\bibinfo {year} {2006})}\BibitemShut {NoStop}%
\bibitem [{\citenamefont {Chu}\ \emph {et~al.}(2016)\citenamefont {Chu},
  \citenamefont {Cao}, \citenamefont {Chen}, \citenamefont {Li},\ and\
  \citenamefont {Zehnder}}]{chu2016experimental}%
  \BibitemOpen
  \bibfield  {author} {\bibinfo {author} {\bibfnamefont {R.}~\bibnamefont
  {Chu}}, \bibinfo {author} {\bibfnamefont {Y.}~\bibnamefont {Cao}}, \bibinfo
  {author} {\bibfnamefont {M.}~\bibnamefont {Chen}}, \bibinfo {author}
  {\bibfnamefont {R.}~\bibnamefont {Li}},\ and\ \bibinfo {author}
  {\bibfnamefont {D.}~\bibnamefont {Zehnder}},\ }\bibfield  {title} {\bibinfo
  {title} {An experimental demonstration of {G}a{N} {CMOS} technology},\
  }\href@noop {} {\bibfield  {journal} {\bibinfo  {journal} {IEEE Electron
  Device Letters}\ }\textbf {\bibinfo {volume} {37}},\ \bibinfo {pages} {269}
  (\bibinfo {year} {2016})}\BibitemShut {NoStop}%
\bibitem [{\citenamefont {Geim}(2009)}]{geim2009graphene}%
  \BibitemOpen
  \bibfield  {author} {\bibinfo {author} {\bibfnamefont {A.~K.}\ \bibnamefont
  {Geim}},\ }\bibfield  {title} {\bibinfo {title} {Graphene: status and
  prospects},\ }\href@noop {} {\bibfield  {journal} {\bibinfo  {journal}
  {science}\ }\textbf {\bibinfo {volume} {324}},\ \bibinfo {pages} {1530}
  (\bibinfo {year} {2009})}\BibitemShut {NoStop}%
\bibitem [{\citenamefont {Neto}\ \emph {et~al.}(2009)\citenamefont {Neto},
  \citenamefont {Guinea}, \citenamefont {Peres}, \citenamefont {Novoselov},\
  and\ \citenamefont {Geim}}]{neto2009electronic}%
  \BibitemOpen
  \bibfield  {author} {\bibinfo {author} {\bibfnamefont {A.~C.}\ \bibnamefont
  {Neto}}, \bibinfo {author} {\bibfnamefont {F.}~\bibnamefont {Guinea}},
  \bibinfo {author} {\bibfnamefont {N.~M.}\ \bibnamefont {Peres}}, \bibinfo
  {author} {\bibfnamefont {K.~S.}\ \bibnamefont {Novoselov}},\ and\ \bibinfo
  {author} {\bibfnamefont {A.~K.}\ \bibnamefont {Geim}},\ }\bibfield  {title}
  {\bibinfo {title} {The electronic properties of graphene},\ }\href@noop {}
  {\bibfield  {journal} {\bibinfo  {journal} {Reviews of Modern Physics}\
  }\textbf {\bibinfo {volume} {81}},\ \bibinfo {pages} {109} (\bibinfo {year}
  {2009})}\BibitemShut {NoStop}%
\bibitem [{\citenamefont {Li}\ \emph {et~al.}(2014)\citenamefont {Li},
  \citenamefont {Yu}, \citenamefont {Ye}, \citenamefont {Ge}, \citenamefont
  {Ou}, \citenamefont {Wu}, \citenamefont {Feng}, \citenamefont {Chen},\ and\
  \citenamefont {Zhang}}]{li2014black}%
  \BibitemOpen
  \bibfield  {author} {\bibinfo {author} {\bibfnamefont {L.}~\bibnamefont
  {Li}}, \bibinfo {author} {\bibfnamefont {Y.}~\bibnamefont {Yu}}, \bibinfo
  {author} {\bibfnamefont {G.~J.}\ \bibnamefont {Ye}}, \bibinfo {author}
  {\bibfnamefont {Q.}~\bibnamefont {Ge}}, \bibinfo {author} {\bibfnamefont
  {X.}~\bibnamefont {Ou}}, \bibinfo {author} {\bibfnamefont {H.}~\bibnamefont
  {Wu}}, \bibinfo {author} {\bibfnamefont {D.}~\bibnamefont {Feng}}, \bibinfo
  {author} {\bibfnamefont {X.~H.}\ \bibnamefont {Chen}},\ and\ \bibinfo
  {author} {\bibfnamefont {Y.}~\bibnamefont {Zhang}},\ }\bibfield  {title}
  {\bibinfo {title} {Black phosphorus field-effect transistors},\ }\href@noop
  {} {\bibfield  {journal} {\bibinfo  {journal} {Nature nanotechnology}\
  }\textbf {\bibinfo {volume} {9}},\ \bibinfo {pages} {372} (\bibinfo {year}
  {2014})}\BibitemShut {NoStop}%
\bibitem [{\citenamefont {Chuang}\ \emph {et~al.}(2014)\citenamefont {Chuang},
  \citenamefont {Battaglia}, \citenamefont {Azcatl}, \citenamefont {McDonnell},
  \citenamefont {Kang}, \citenamefont {Yin}, \citenamefont {Tosun},
  \citenamefont {Kapadia}, \citenamefont {Fang}, \citenamefont {Wallace} \emph
  {et~al.}}]{chuang2014mos2}%
  \BibitemOpen
  \bibfield  {author} {\bibinfo {author} {\bibfnamefont {S.}~\bibnamefont
  {Chuang}}, \bibinfo {author} {\bibfnamefont {C.}~\bibnamefont {Battaglia}},
  \bibinfo {author} {\bibfnamefont {A.}~\bibnamefont {Azcatl}}, \bibinfo
  {author} {\bibfnamefont {S.}~\bibnamefont {McDonnell}}, \bibinfo {author}
  {\bibfnamefont {J.~S.}\ \bibnamefont {Kang}}, \bibinfo {author}
  {\bibfnamefont {X.}~\bibnamefont {Yin}}, \bibinfo {author} {\bibfnamefont
  {M.}~\bibnamefont {Tosun}}, \bibinfo {author} {\bibfnamefont
  {R.}~\bibnamefont {Kapadia}}, \bibinfo {author} {\bibfnamefont
  {H.}~\bibnamefont {Fang}}, \bibinfo {author} {\bibfnamefont {R.~M.}\
  \bibnamefont {Wallace}}, \emph {et~al.},\ }\bibfield  {title} {\bibinfo
  {title} {{M}o{S}$_2$ p-type transistors and diodes enabled by high work
  function moo x contacts},\ }\href@noop {} {\bibfield  {journal} {\bibinfo
  {journal} {Nano letters}\ }\textbf {\bibinfo {volume} {14}},\ \bibinfo
  {pages} {1337} (\bibinfo {year} {2014})}\BibitemShut {NoStop}%
\bibitem [{\citenamefont {Zhan}\ \emph {et~al.}(2014)\citenamefont {Zhan},
  \citenamefont {Li}, \citenamefont {Yang}, \citenamefont {Jenkins},
  \citenamefont {Huang},\ and\ \citenamefont {Dong}}]{zhan2014graphene}%
  \BibitemOpen
  \bibfield  {author} {\bibinfo {author} {\bibfnamefont {B.}~\bibnamefont
  {Zhan}}, \bibinfo {author} {\bibfnamefont {C.}~\bibnamefont {Li}}, \bibinfo
  {author} {\bibfnamefont {J.}~\bibnamefont {Yang}}, \bibinfo {author}
  {\bibfnamefont {G.}~\bibnamefont {Jenkins}}, \bibinfo {author} {\bibfnamefont
  {W.}~\bibnamefont {Huang}},\ and\ \bibinfo {author} {\bibfnamefont
  {X.}~\bibnamefont {Dong}},\ }\bibfield  {title} {\bibinfo {title} {Graphene
  field-effect transistor and its application for electronic sensing},\
  }\href@noop {} {\bibfield  {journal} {\bibinfo  {journal} {Small}\ }\textbf
  {\bibinfo {volume} {10}},\ \bibinfo {pages} {4042} (\bibinfo {year}
  {2014})}\BibitemShut {NoStop}%
\bibitem [{\citenamefont {Dean}\ \emph {et~al.}(2012)\citenamefont {Dean},
  \citenamefont {Young}, \citenamefont {Wang}, \citenamefont {Meric},
  \citenamefont {Lee}, \citenamefont {Watanabe}, \citenamefont {Taniguchi},
  \citenamefont {Shepard}, \citenamefont {Kim},\ and\ \citenamefont
  {Hone}}]{dean2012graphene}%
  \BibitemOpen
  \bibfield  {author} {\bibinfo {author} {\bibfnamefont {C.}~\bibnamefont
  {Dean}}, \bibinfo {author} {\bibfnamefont {A.}~\bibnamefont {Young}},
  \bibinfo {author} {\bibfnamefont {L.}~\bibnamefont {Wang}}, \bibinfo {author}
  {\bibfnamefont {I.}~\bibnamefont {Meric}}, \bibinfo {author} {\bibfnamefont
  {G.-H.}\ \bibnamefont {Lee}}, \bibinfo {author} {\bibfnamefont
  {K.}~\bibnamefont {Watanabe}}, \bibinfo {author} {\bibfnamefont
  {T.}~\bibnamefont {Taniguchi}}, \bibinfo {author} {\bibfnamefont
  {K.}~\bibnamefont {Shepard}}, \bibinfo {author} {\bibfnamefont
  {P.}~\bibnamefont {Kim}},\ and\ \bibinfo {author} {\bibfnamefont
  {J.}~\bibnamefont {Hone}},\ }\bibfield  {title} {\bibinfo {title} {Graphene
  based heterostructures},\ }\href@noop {} {\bibfield  {journal} {\bibinfo
  {journal} {Solid State Communications}\ }\textbf {\bibinfo {volume} {152}},\
  \bibinfo {pages} {1275} (\bibinfo {year} {2012})}\BibitemShut {NoStop}%
\bibitem [{\citenamefont {Wang}\ \emph {et~al.}(2019)\citenamefont {Wang},
  \citenamefont {Ren}, \citenamefont {Yan}, \citenamefont {Jiang},
  \citenamefont {Sha},\ and\ \citenamefont {Shan}}]{wang2019graphene}%
  \BibitemOpen
  \bibfield  {author} {\bibinfo {author} {\bibfnamefont {R.}~\bibnamefont
  {Wang}}, \bibinfo {author} {\bibfnamefont {X.-G.}\ \bibnamefont {Ren}},
  \bibinfo {author} {\bibfnamefont {Z.}~\bibnamefont {Yan}}, \bibinfo {author}
  {\bibfnamefont {L.-J.}\ \bibnamefont {Jiang}}, \bibinfo {author}
  {\bibfnamefont {W.~E.}\ \bibnamefont {Sha}},\ and\ \bibinfo {author}
  {\bibfnamefont {G.-C.}\ \bibnamefont {Shan}},\ }\bibfield  {title} {\bibinfo
  {title} {Graphene based functional devices: A short review},\ }\href@noop {}
  {\bibfield  {journal} {\bibinfo  {journal} {Frontiers of Physics}\ }\textbf
  {\bibinfo {volume} {14}},\ \bibinfo {pages} {1} (\bibinfo {year}
  {2019})}\BibitemShut {NoStop}%
\bibitem [{\citenamefont {Wang}\ \emph {et~al.}(2012)\citenamefont {Wang},
  \citenamefont {Kalantar-Zadeh}, \citenamefont {Kis}, \citenamefont
  {Coleman},\ and\ \citenamefont {Strano}}]{wang2012electronics}%
  \BibitemOpen
  \bibfield  {author} {\bibinfo {author} {\bibfnamefont {Q.~H.}\ \bibnamefont
  {Wang}}, \bibinfo {author} {\bibfnamefont {K.}~\bibnamefont
  {Kalantar-Zadeh}}, \bibinfo {author} {\bibfnamefont {A.}~\bibnamefont {Kis}},
  \bibinfo {author} {\bibfnamefont {J.~N.}\ \bibnamefont {Coleman}},\ and\
  \bibinfo {author} {\bibfnamefont {M.~S.}\ \bibnamefont {Strano}},\ }\bibfield
   {title} {\bibinfo {title} {Electronics and optoelectronics of
  two-dimensional transition metal dichalcogenides},\ }\href@noop {} {\bibfield
   {journal} {\bibinfo  {journal} {Nature nanotechnology}\ }\textbf {\bibinfo
  {volume} {7}},\ \bibinfo {pages} {699} (\bibinfo {year} {2012})}\BibitemShut
  {NoStop}%
\bibitem [{\citenamefont {Lee}\ \emph {et~al.}(2014)\citenamefont {Lee},
  \citenamefont {Lee}, \citenamefont {Van Der~Zande}, \citenamefont {Chen},
  \citenamefont {Li}, \citenamefont {Han}, \citenamefont {Cui}, \citenamefont
  {Arefe}, \citenamefont {Nuckolls}, \citenamefont {Heinz} \emph
  {et~al.}}]{lee2014atomically}%
  \BibitemOpen
  \bibfield  {author} {\bibinfo {author} {\bibfnamefont {C.-H.}\ \bibnamefont
  {Lee}}, \bibinfo {author} {\bibfnamefont {G.-H.}\ \bibnamefont {Lee}},
  \bibinfo {author} {\bibfnamefont {A.~M.}\ \bibnamefont {Van Der~Zande}},
  \bibinfo {author} {\bibfnamefont {W.}~\bibnamefont {Chen}}, \bibinfo {author}
  {\bibfnamefont {Y.}~\bibnamefont {Li}}, \bibinfo {author} {\bibfnamefont
  {M.}~\bibnamefont {Han}}, \bibinfo {author} {\bibfnamefont {X.}~\bibnamefont
  {Cui}}, \bibinfo {author} {\bibfnamefont {G.}~\bibnamefont {Arefe}}, \bibinfo
  {author} {\bibfnamefont {C.}~\bibnamefont {Nuckolls}}, \bibinfo {author}
  {\bibfnamefont {T.~F.}\ \bibnamefont {Heinz}}, \emph {et~al.},\ }\bibfield
  {title} {\bibinfo {title} {Atomically thin p--n junctions with van der
  {W}aals heterointerfaces},\ }\href@noop {} {\bibfield  {journal} {\bibinfo
  {journal} {Nature nanotechnology}\ }\textbf {\bibinfo {volume} {9}},\
  \bibinfo {pages} {676} (\bibinfo {year} {2014})}\BibitemShut {NoStop}%
\bibitem [{\citenamefont {Ashley}\ \emph {et~al.}(2007)\citenamefont {Ashley},
  \citenamefont {Buckle}, \citenamefont {Datta}, \citenamefont {Emeny},
  \citenamefont {Hayes}, \citenamefont {Hilton}, \citenamefont {Jefferies},
  \citenamefont {Martin}, \citenamefont {Phillips}, \citenamefont {Wallis}
  \emph {et~al.}}]{ashley2007heterogeneous}%
  \BibitemOpen
  \bibfield  {author} {\bibinfo {author} {\bibfnamefont {T.}~\bibnamefont
  {Ashley}}, \bibinfo {author} {\bibfnamefont {L.}~\bibnamefont {Buckle}},
  \bibinfo {author} {\bibfnamefont {S.}~\bibnamefont {Datta}}, \bibinfo
  {author} {\bibfnamefont {M.}~\bibnamefont {Emeny}}, \bibinfo {author}
  {\bibfnamefont {D.}~\bibnamefont {Hayes}}, \bibinfo {author} {\bibfnamefont
  {K.}~\bibnamefont {Hilton}}, \bibinfo {author} {\bibfnamefont
  {R.}~\bibnamefont {Jefferies}}, \bibinfo {author} {\bibfnamefont
  {T.}~\bibnamefont {Martin}}, \bibinfo {author} {\bibfnamefont
  {T.}~\bibnamefont {Phillips}}, \bibinfo {author} {\bibfnamefont
  {D.}~\bibnamefont {Wallis}}, \emph {et~al.},\ }\bibfield  {title} {\bibinfo
  {title} {Heterogeneous {I}n{S}b quantum well transistors on silicon for
  ultra-high speed, low power logic applications},\ }\href@noop {} {\bibfield
  {journal} {\bibinfo  {journal} {Electronics Letters}\ }\textbf {\bibinfo
  {volume} {43}},\ \bibinfo {pages} {777} (\bibinfo {year} {2007})}\BibitemShut
  {NoStop}%
\bibitem [{\citenamefont {Doornbos}\ and\ \citenamefont
  {Passlack}(2010)}]{Doornbos2010}%
  \BibitemOpen
  \bibfield  {author} {\bibinfo {author} {\bibfnamefont {G.}~\bibnamefont
  {Doornbos}}\ and\ \bibinfo {author} {\bibfnamefont {M.}~\bibnamefont
  {Passlack}},\ }\bibfield  {title} {\bibinfo {title} {Benchmarking of iii–v
  n-{MOSFET} maturity and feasibility for future cmos},\ }\href
  {https://doi.org/10.1109/LED.2010.2063012} {\bibfield  {journal} {\bibinfo
  {journal} {IEEE Electron Device Letters}\ }\textbf {\bibinfo {volume} {31}},\
  \bibinfo {pages} {1110} (\bibinfo {year} {2010})}\BibitemShut {NoStop}%
\bibitem [{\citenamefont {Vurgaftman}\ \emph {et~al.}(2001)\citenamefont
  {Vurgaftman}, \citenamefont {Meyer},\ and\ \citenamefont
  {Ram-Mohan}}]{vurgaftman2001band}%
  \BibitemOpen
  \bibfield  {author} {\bibinfo {author} {\bibfnamefont {I.}~\bibnamefont
  {Vurgaftman}}, \bibinfo {author} {\bibfnamefont {J.~{\'a}.}\ \bibnamefont
  {Meyer}},\ and\ \bibinfo {author} {\bibfnamefont {L.~{\'a}.}\ \bibnamefont
  {Ram-Mohan}},\ }\bibfield  {title} {\bibinfo {title} {Band parameters for
  iii--v compound semiconductors and their alloys},\ }\href@noop {} {\bibfield
  {journal} {\bibinfo  {journal} {Journal of applied physics}\ }\textbf
  {\bibinfo {volume} {89}},\ \bibinfo {pages} {5815} (\bibinfo {year}
  {2001})}\BibitemShut {NoStop}%
\bibitem [{\citenamefont {Kallaher}\ \emph {et~al.}(2010)\citenamefont
  {Kallaher}, \citenamefont {Heremans}, \citenamefont {Goel}, \citenamefont
  {Chung},\ and\ \citenamefont {Santos}}]{kallaher2010spin}%
  \BibitemOpen
  \bibfield  {author} {\bibinfo {author} {\bibfnamefont {R.}~\bibnamefont
  {Kallaher}}, \bibinfo {author} {\bibfnamefont {J.~J.}\ \bibnamefont
  {Heremans}}, \bibinfo {author} {\bibfnamefont {N.}~\bibnamefont {Goel}},
  \bibinfo {author} {\bibfnamefont {S.}~\bibnamefont {Chung}},\ and\ \bibinfo
  {author} {\bibfnamefont {M.}~\bibnamefont {Santos}},\ }\bibfield  {title}
  {\bibinfo {title} {Spin-orbit interaction determined by antilocalization in
  an {I}n{S}b quantum well},\ }\href@noop {} {\bibfield  {journal} {\bibinfo
  {journal} {Physical Review B}\ }\textbf {\bibinfo {volume} {81}},\ \bibinfo
  {pages} {075303} (\bibinfo {year} {2010})}\BibitemShut {NoStop}%
\bibitem [{\citenamefont {Khodaparast}\ \emph {et~al.}(2004)\citenamefont
  {Khodaparast}, \citenamefont {Doezema}, \citenamefont {Chung}, \citenamefont
  {Goldammer},\ and\ \citenamefont {Santos}}]{khodaparast2004spectroscopy}%
  \BibitemOpen
  \bibfield  {author} {\bibinfo {author} {\bibfnamefont {G.~A.}\ \bibnamefont
  {Khodaparast}}, \bibinfo {author} {\bibfnamefont {R.}~\bibnamefont
  {Doezema}}, \bibinfo {author} {\bibfnamefont {S.}~\bibnamefont {Chung}},
  \bibinfo {author} {\bibfnamefont {K.}~\bibnamefont {Goldammer}},\ and\
  \bibinfo {author} {\bibfnamefont {M.}~\bibnamefont {Santos}},\ }\bibfield
  {title} {\bibinfo {title} {Spectroscopy of rashba spin splitting in {I}n{S}b
  quantum wells},\ }\href@noop {} {\bibfield  {journal} {\bibinfo  {journal}
  {Physical Review B}\ }\textbf {\bibinfo {volume} {70}},\ \bibinfo {pages}
  {155322} (\bibinfo {year} {2004})}\BibitemShut {NoStop}%
\bibitem [{\citenamefont {Leontiadou}\ \emph {et~al.}(2011)\citenamefont
  {Leontiadou}, \citenamefont {Litvinenko}, \citenamefont {Gilbertson},
  \citenamefont {Pidgeon}, \citenamefont {Branford}, \citenamefont {Cohen},
  \citenamefont {Fearn}, \citenamefont {Ashley}, \citenamefont {Emeny},
  \citenamefont {Murdin} \emph {et~al.}}]{leontiadou2011experimental}%
  \BibitemOpen
  \bibfield  {author} {\bibinfo {author} {\bibfnamefont {M.}~\bibnamefont
  {Leontiadou}}, \bibinfo {author} {\bibfnamefont {K.}~\bibnamefont
  {Litvinenko}}, \bibinfo {author} {\bibfnamefont {A.}~\bibnamefont
  {Gilbertson}}, \bibinfo {author} {\bibfnamefont {C.}~\bibnamefont {Pidgeon}},
  \bibinfo {author} {\bibfnamefont {W.}~\bibnamefont {Branford}}, \bibinfo
  {author} {\bibfnamefont {L.}~\bibnamefont {Cohen}}, \bibinfo {author}
  {\bibfnamefont {M.}~\bibnamefont {Fearn}}, \bibinfo {author} {\bibfnamefont
  {T.}~\bibnamefont {Ashley}}, \bibinfo {author} {\bibfnamefont
  {M.}~\bibnamefont {Emeny}}, \bibinfo {author} {\bibfnamefont
  {B.}~\bibnamefont {Murdin}}, \emph {et~al.},\ }\bibfield  {title} {\bibinfo
  {title} {Experimental determination of the rashba coefficient in
  {I}n{S}b/{I}n{A}l{S}b quantum wells at zero magnetic field and elevated
  temperatures},\ }\href@noop {} {\bibfield  {journal} {\bibinfo  {journal}
  {Journal of Physics: Condensed Matter}\ }\textbf {\bibinfo {volume} {23}},\
  \bibinfo {pages} {035801} (\bibinfo {year} {2011})}\BibitemShut {NoStop}%
\bibitem [{\citenamefont {Gilbertson}\ \emph {et~al.}(2009)\citenamefont
  {Gilbertson}, \citenamefont {Branford}, \citenamefont {Fearn}, \citenamefont
  {Buckle}, \citenamefont {Buckle}, \citenamefont {Ashley},\ and\ \citenamefont
  {Cohen}}]{gilbertson2009zero}%
  \BibitemOpen
  \bibfield  {author} {\bibinfo {author} {\bibfnamefont {A.}~\bibnamefont
  {Gilbertson}}, \bibinfo {author} {\bibfnamefont {W.}~\bibnamefont
  {Branford}}, \bibinfo {author} {\bibfnamefont {M.}~\bibnamefont {Fearn}},
  \bibinfo {author} {\bibfnamefont {L.}~\bibnamefont {Buckle}}, \bibinfo
  {author} {\bibfnamefont {P.~D.}\ \bibnamefont {Buckle}}, \bibinfo {author}
  {\bibfnamefont {T.}~\bibnamefont {Ashley}},\ and\ \bibinfo {author}
  {\bibfnamefont {L.}~\bibnamefont {Cohen}},\ }\bibfield  {title} {\bibinfo
  {title} {Zero-field spin splitting and spin-dependent broadening in
  high-mobility {I}n{S}b/in 1- x al x sb asymmetric quantum well
  heterostructures},\ }\href@noop {} {\bibfield  {journal} {\bibinfo  {journal}
  {Physical Review B}\ }\textbf {\bibinfo {volume} {79}},\ \bibinfo {pages}
  {235333} (\bibinfo {year} {2009})}\BibitemShut {NoStop}%
\bibitem [{\citenamefont {Orr}\ \emph {et~al.}(2007)\citenamefont {Orr},
  \citenamefont {Buckle}, \citenamefont {Fearn}, \citenamefont {Storey},
  \citenamefont {Buckle},\ and\ \citenamefont {Ashley}}]{orr2007surface}%
  \BibitemOpen
  \bibfield  {author} {\bibinfo {author} {\bibfnamefont {J.}~\bibnamefont
  {Orr}}, \bibinfo {author} {\bibfnamefont {P.~D.}\ \bibnamefont {Buckle}},
  \bibinfo {author} {\bibfnamefont {M.}~\bibnamefont {Fearn}}, \bibinfo
  {author} {\bibfnamefont {C.}~\bibnamefont {Storey}}, \bibinfo {author}
  {\bibfnamefont {L.}~\bibnamefont {Buckle}},\ and\ \bibinfo {author}
  {\bibfnamefont {T.}~\bibnamefont {Ashley}},\ }\bibfield  {title} {\bibinfo
  {title} {A surface-gated {I}n{S}b quantum well single electron transistor},\
  }\href@noop {} {\bibfield  {journal} {\bibinfo  {journal} {New Journal of
  Physics}\ }\textbf {\bibinfo {volume} {9}},\ \bibinfo {pages} {261} (\bibinfo
  {year} {2007})}\BibitemShut {NoStop}%
\bibitem [{\citenamefont {Madelung}\ \emph {et~al.}(2000)\citenamefont
  {Madelung}, \citenamefont {Roessler},\ and\ \citenamefont
  {Schulz}}]{Madelung}%
  \BibitemOpen
  \bibfield  {author} {\bibinfo {author} {\bibfnamefont {O.}~\bibnamefont
  {Madelung}}, \bibinfo {author} {\bibfnamefont {U.}~\bibnamefont {Roessler}},\
  and\ \bibinfo {author} {\bibfnamefont {M.}~\bibnamefont {Schulz}},\
  }\href@noop {} {\emph {\bibinfo {title} {Landolt-BernsteinGroup III Condensed
  Matter, 41D}}}\ (\bibinfo  {publisher} {Springer, Berlin},\ \bibinfo {year}
  {2000})\BibitemShut {NoStop}%
\bibitem [{\citenamefont {Ashley}\ \emph {et~al.}(1995)\citenamefont {Ashley},
  \citenamefont {Dean}, \citenamefont {Elliott}, \citenamefont {Pryce},
  \citenamefont {Johnson},\ and\ \citenamefont {Willis}}]{Ashley1995}%
  \BibitemOpen
  \bibfield  {author} {\bibinfo {author} {\bibfnamefont {T.}~\bibnamefont
  {Ashley}}, \bibinfo {author} {\bibfnamefont {A.~B.}\ \bibnamefont {Dean}},
  \bibinfo {author} {\bibfnamefont {C.~T.}\ \bibnamefont {Elliott}}, \bibinfo
  {author} {\bibfnamefont {G.~J.}\ \bibnamefont {Pryce}}, \bibinfo {author}
  {\bibfnamefont {A.~D.}\ \bibnamefont {Johnson}},\ and\ \bibinfo {author}
  {\bibfnamefont {H.}~\bibnamefont {Willis}},\ }\bibfield  {title} {\bibinfo
  {title} {Uncooled high-speed {I}n{S}b field-effect transistors},\ }\href
  {https://doi.org/10.1063/1.114063} {\bibfield  {journal} {\bibinfo  {journal}
  {Applied Physics Letters}\ }\textbf {\bibinfo {volume} {66}},\ \bibinfo
  {pages} {481} (\bibinfo {year} {1995})}\BibitemShut {NoStop}%
\bibitem [{\citenamefont {Chen}\ \emph {et~al.}(2005)\citenamefont {Chen},
  \citenamefont {Heremans}, \citenamefont {Peters}, \citenamefont {Govorov},
  \citenamefont {Goel}, \citenamefont {Chung},\ and\ \citenamefont
  {Santos}}]{chen2005spin}%
  \BibitemOpen
  \bibfield  {author} {\bibinfo {author} {\bibfnamefont {H.}~\bibnamefont
  {Chen}}, \bibinfo {author} {\bibfnamefont {J.}~\bibnamefont {Heremans}},
  \bibinfo {author} {\bibfnamefont {J.}~\bibnamefont {Peters}}, \bibinfo
  {author} {\bibfnamefont {A.}~\bibnamefont {Govorov}}, \bibinfo {author}
  {\bibfnamefont {N.}~\bibnamefont {Goel}}, \bibinfo {author} {\bibfnamefont
  {S.}~\bibnamefont {Chung}},\ and\ \bibinfo {author} {\bibfnamefont
  {M.}~\bibnamefont {Santos}},\ }\bibfield  {title} {\bibinfo {title}
  {Spin-polarized reflection in a two-dimensional electron system},\
  }\href@noop {} {\bibfield  {journal} {\bibinfo  {journal} {Applied Physics
  Letters}\ }\textbf {\bibinfo {volume} {86}},\ \bibinfo {pages} {032113}
  (\bibinfo {year} {2005})}\BibitemShut {NoStop}%
\bibitem [{\citenamefont {{\v{Z}}uti{\'c}}\ \emph {et~al.}(2004)\citenamefont
  {{\v{Z}}uti{\'c}}, \citenamefont {Fabian},\ and\ \citenamefont
  {Sarma}}]{vzutic2004spintronics}%
  \BibitemOpen
  \bibfield  {author} {\bibinfo {author} {\bibfnamefont {I.}~\bibnamefont
  {{\v{Z}}uti{\'c}}}, \bibinfo {author} {\bibfnamefont {J.}~\bibnamefont
  {Fabian}},\ and\ \bibinfo {author} {\bibfnamefont {S.~D.}\ \bibnamefont
  {Sarma}},\ }\bibfield  {title} {\bibinfo {title} {Spintronics: Fundamentals
  and applications},\ }\href@noop {} {\bibfield  {journal} {\bibinfo  {journal}
  {Reviews of Modern Physics}\ }\textbf {\bibinfo {volume} {76}},\ \bibinfo
  {pages} {323} (\bibinfo {year} {2004})}\BibitemShut {NoStop}%
\bibitem [{\citenamefont {Oreg}\ \emph {et~al.}(2010)\citenamefont {Oreg},
  \citenamefont {Refael},\ and\ \citenamefont {Von~Oppen}}]{oreg2010helical}%
  \BibitemOpen
  \bibfield  {author} {\bibinfo {author} {\bibfnamefont {Y.}~\bibnamefont
  {Oreg}}, \bibinfo {author} {\bibfnamefont {G.}~\bibnamefont {Refael}},\ and\
  \bibinfo {author} {\bibfnamefont {F.}~\bibnamefont {Von~Oppen}},\ }\bibfield
  {title} {\bibinfo {title} {Helical liquids and majorana bound states in
  quantum wires},\ }\href@noop {} {\bibfield  {journal} {\bibinfo  {journal}
  {Physical review letters}\ }\textbf {\bibinfo {volume} {105}},\ \bibinfo
  {pages} {177002} (\bibinfo {year} {2010})}\BibitemShut {NoStop}%
\bibitem [{\citenamefont {Ke}\ \emph {et~al.}(2019)\citenamefont {Ke},
  \citenamefont {Moehle}, \citenamefont {de~Vries}, \citenamefont {Thomas},
  \citenamefont {Metti}, \citenamefont {Guinn}, \citenamefont {Kallaher},
  \citenamefont {Lodari}, \citenamefont {Scappucci}, \citenamefont {Wang} \emph
  {et~al.}}]{ke2019ballistic}%
  \BibitemOpen
  \bibfield  {author} {\bibinfo {author} {\bibfnamefont {C.~T.}\ \bibnamefont
  {Ke}}, \bibinfo {author} {\bibfnamefont {C.~M.}\ \bibnamefont {Moehle}},
  \bibinfo {author} {\bibfnamefont {F.~K.}\ \bibnamefont {de~Vries}}, \bibinfo
  {author} {\bibfnamefont {C.}~\bibnamefont {Thomas}}, \bibinfo {author}
  {\bibfnamefont {S.}~\bibnamefont {Metti}}, \bibinfo {author} {\bibfnamefont
  {C.~R.}\ \bibnamefont {Guinn}}, \bibinfo {author} {\bibfnamefont
  {R.}~\bibnamefont {Kallaher}}, \bibinfo {author} {\bibfnamefont
  {M.}~\bibnamefont {Lodari}}, \bibinfo {author} {\bibfnamefont
  {G.}~\bibnamefont {Scappucci}}, \bibinfo {author} {\bibfnamefont
  {T.}~\bibnamefont {Wang}}, \emph {et~al.},\ }\bibfield  {title} {\bibinfo
  {title} {Ballistic superconductivity and tunable $\pi$--junctions in {I}n{S}b
  quantum wells},\ }\href@noop {} {\bibfield  {journal} {\bibinfo  {journal}
  {Nature communications}\ }\textbf {\bibinfo {volume} {10}},\ \bibinfo {pages}
  {1} (\bibinfo {year} {2019})}\BibitemShut {NoStop}%
\bibitem [{\citenamefont {Deng}\ \emph {et~al.}(2016)\citenamefont {Deng},
  \citenamefont {Vaitiek{\.e}nas}, \citenamefont {Hansen}, \citenamefont
  {Danon}, \citenamefont {Leijnse}, \citenamefont {Flensberg}, \citenamefont
  {Nyg{\aa}rd}, \citenamefont {Krogstrup},\ and\ \citenamefont
  {Marcus}}]{deng2016majorana}%
  \BibitemOpen
  \bibfield  {author} {\bibinfo {author} {\bibfnamefont {M.}~\bibnamefont
  {Deng}}, \bibinfo {author} {\bibfnamefont {S.}~\bibnamefont
  {Vaitiek{\.e}nas}}, \bibinfo {author} {\bibfnamefont {E.~B.}\ \bibnamefont
  {Hansen}}, \bibinfo {author} {\bibfnamefont {J.}~\bibnamefont {Danon}},
  \bibinfo {author} {\bibfnamefont {M.}~\bibnamefont {Leijnse}}, \bibinfo
  {author} {\bibfnamefont {K.}~\bibnamefont {Flensberg}}, \bibinfo {author}
  {\bibfnamefont {J.}~\bibnamefont {Nyg{\aa}rd}}, \bibinfo {author}
  {\bibfnamefont {P.}~\bibnamefont {Krogstrup}},\ and\ \bibinfo {author}
  {\bibfnamefont {C.~M.}\ \bibnamefont {Marcus}},\ }\bibfield  {title}
  {\bibinfo {title} {Majorana bound state in a coupled quantum-dot
  hybrid-nanowire system},\ }\href@noop {} {\bibfield  {journal} {\bibinfo
  {journal} {Science}\ }\textbf {\bibinfo {volume} {354}},\ \bibinfo {pages}
  {1557} (\bibinfo {year} {2016})}\BibitemShut {NoStop}%
\bibitem [{\citenamefont {Lehner}\ \emph {et~al.}(2018)\citenamefont {Lehner},
  \citenamefont {Tschirky}, \citenamefont {Ihn}, \citenamefont {Dietsche},
  \citenamefont {Keller}, \citenamefont {F\"alt},\ and\ \citenamefont
  {Wegscheider}}]{Lehner2018}%
  \BibitemOpen
  \bibfield  {author} {\bibinfo {author} {\bibfnamefont {C.~A.}\ \bibnamefont
  {Lehner}}, \bibinfo {author} {\bibfnamefont {T.}~\bibnamefont {Tschirky}},
  \bibinfo {author} {\bibfnamefont {T.}~\bibnamefont {Ihn}}, \bibinfo {author}
  {\bibfnamefont {W.}~\bibnamefont {Dietsche}}, \bibinfo {author}
  {\bibfnamefont {J.}~\bibnamefont {Keller}}, \bibinfo {author} {\bibfnamefont
  {S.}~\bibnamefont {F\"alt}},\ and\ \bibinfo {author} {\bibfnamefont
  {W.}~\bibnamefont {Wegscheider}},\ }\bibfield  {title} {\bibinfo {title}
  {Limiting scattering processes in high-mobility {I}n{S}b quantum wells grown
  on gasb buffer systems},\ }\href
  {https://doi.org/10.1103/PhysRevMaterials.2.054601} {\bibfield  {journal}
  {\bibinfo  {journal} {Physical Review Materials}\ }\textbf {\bibinfo {volume}
  {2}},\ \bibinfo {pages} {054601} (\bibinfo {year} {2018})}\BibitemShut
  {NoStop}%
\bibitem [{\citenamefont {Lei}\ \emph {et~al.}(2019)\citenamefont {Lei},
  \citenamefont {Lehner}, \citenamefont {Cheah}, \citenamefont {Karalic},
  \citenamefont {Mittag}, \citenamefont {Alt}, \citenamefont {Scharnetzky},
  \citenamefont {Wegscheider}, \citenamefont {Ihn},\ and\ \citenamefont
  {Ensslin}}]{lei2019quantum}%
  \BibitemOpen
  \bibfield  {author} {\bibinfo {author} {\bibfnamefont {Z.}~\bibnamefont
  {Lei}}, \bibinfo {author} {\bibfnamefont {C.~A.}\ \bibnamefont {Lehner}},
  \bibinfo {author} {\bibfnamefont {E.}~\bibnamefont {Cheah}}, \bibinfo
  {author} {\bibfnamefont {M.}~\bibnamefont {Karalic}}, \bibinfo {author}
  {\bibfnamefont {C.}~\bibnamefont {Mittag}}, \bibinfo {author} {\bibfnamefont
  {L.}~\bibnamefont {Alt}}, \bibinfo {author} {\bibfnamefont {J.}~\bibnamefont
  {Scharnetzky}}, \bibinfo {author} {\bibfnamefont {W.}~\bibnamefont
  {Wegscheider}}, \bibinfo {author} {\bibfnamefont {T.}~\bibnamefont {Ihn}},\
  and\ \bibinfo {author} {\bibfnamefont {K.}~\bibnamefont {Ensslin}},\
  }\bibfield  {title} {\bibinfo {title} {Quantum transport in high-quality
  shallow {I}n{S}b quantum wells},\ }\href@noop {} {\bibfield  {journal}
  {\bibinfo  {journal} {Applied Physics Letters}\ }\textbf {\bibinfo {volume}
  {115}},\ \bibinfo {pages} {012101} (\bibinfo {year} {2019})}\BibitemShut
  {NoStop}%
\bibitem [{\citenamefont {Nilsson}\ \emph {et~al.}(2009)\citenamefont
  {Nilsson}, \citenamefont {Caroff}, \citenamefont {Thelander}, \citenamefont
  {Larsson}, \citenamefont {Wagner}, \citenamefont {Wernersson}, \citenamefont
  {Samuelson},\ and\ \citenamefont {Xu}}]{Nilsson2009}%
  \BibitemOpen
  \bibfield  {author} {\bibinfo {author} {\bibfnamefont {H.~A.}\ \bibnamefont
  {Nilsson}}, \bibinfo {author} {\bibfnamefont {P.}~\bibnamefont {Caroff}},
  \bibinfo {author} {\bibfnamefont {C.}~\bibnamefont {Thelander}}, \bibinfo
  {author} {\bibfnamefont {M.}~\bibnamefont {Larsson}}, \bibinfo {author}
  {\bibfnamefont {J.~B.}\ \bibnamefont {Wagner}}, \bibinfo {author}
  {\bibfnamefont {L.-E.}\ \bibnamefont {Wernersson}}, \bibinfo {author}
  {\bibfnamefont {L.}~\bibnamefont {Samuelson}},\ and\ \bibinfo {author}
  {\bibfnamefont {H.~Q.}\ \bibnamefont {Xu}},\ }\bibfield  {title} {\bibinfo
  {title} {Giant, level-dependent g factors in {I}n{S}b nanowire quantum
  dots},\ }\href {https://doi.org/10.1021/nl901333a} {\bibfield  {journal}
  {\bibinfo  {journal} {Nano Letters}\ }\textbf {\bibinfo {volume} {9}},\
  \bibinfo {pages} {3151} (\bibinfo {year} {2009})},\ \bibinfo {note} {pMID:
  19736971},\ \Eprint {https://arxiv.org/abs/https://doi.org/10.1021/nl901333a}
  {https://doi.org/10.1021/nl901333a} \BibitemShut {NoStop}%
\bibitem [{\citenamefont {Nilsson}\ \emph {et~al.}(2010)\citenamefont
  {Nilsson}, \citenamefont {Karlstr\"om}, \citenamefont {Larsson},
  \citenamefont {Caroff}, \citenamefont {Pedersen}, \citenamefont {Samuelson},
  \citenamefont {Wacker}, \citenamefont {Wernersson},\ and\ \citenamefont
  {Xu}}]{Nilsson2010}%
  \BibitemOpen
  \bibfield  {author} {\bibinfo {author} {\bibfnamefont {H.~A.}\ \bibnamefont
  {Nilsson}}, \bibinfo {author} {\bibfnamefont {O.}~\bibnamefont
  {Karlstr\"om}}, \bibinfo {author} {\bibfnamefont {M.}~\bibnamefont
  {Larsson}}, \bibinfo {author} {\bibfnamefont {P.}~\bibnamefont {Caroff}},
  \bibinfo {author} {\bibfnamefont {J.~N.}\ \bibnamefont {Pedersen}}, \bibinfo
  {author} {\bibfnamefont {L.}~\bibnamefont {Samuelson}}, \bibinfo {author}
  {\bibfnamefont {A.}~\bibnamefont {Wacker}}, \bibinfo {author} {\bibfnamefont
  {L.-E.}\ \bibnamefont {Wernersson}},\ and\ \bibinfo {author} {\bibfnamefont
  {H.~Q.}\ \bibnamefont {Xu}},\ }\bibfield  {title} {\bibinfo {title}
  {Correlation-induced conductance suppression at level degeneracy in a quantum
  dot},\ }\href {https://doi.org/10.1103/PhysRevLett.104.186804} {\bibfield
  {journal} {\bibinfo  {journal} {Physical Review Letters}\ }\textbf {\bibinfo
  {volume} {104}},\ \bibinfo {pages} {186804} (\bibinfo {year}
  {2010})}\BibitemShut {NoStop}%
\bibitem [{\citenamefont {De~La~Mata}\ \emph {et~al.}(2016)\citenamefont
  {De~La~Mata}, \citenamefont {Leturcq}, \citenamefont {Plissard},
  \citenamefont {Rolland}, \citenamefont {Mag{\'e}n}, \citenamefont {Arbiol},\
  and\ \citenamefont {Caroff}}]{Mata2016twin}%
  \BibitemOpen
  \bibfield  {author} {\bibinfo {author} {\bibfnamefont {M.}~\bibnamefont
  {De~La~Mata}}, \bibinfo {author} {\bibfnamefont {R.}~\bibnamefont {Leturcq}},
  \bibinfo {author} {\bibfnamefont {S.~R.}\ \bibnamefont {Plissard}}, \bibinfo
  {author} {\bibfnamefont {C.}~\bibnamefont {Rolland}}, \bibinfo {author}
  {\bibfnamefont {C.}~\bibnamefont {Mag{\'e}n}}, \bibinfo {author}
  {\bibfnamefont {J.}~\bibnamefont {Arbiol}},\ and\ \bibinfo {author}
  {\bibfnamefont {P.}~\bibnamefont {Caroff}},\ }\bibfield  {title} {\bibinfo
  {title} {Twin-induced {I}n{S}b nanosails: a convenient high mobility quantum
  system},\ }\href@noop {} {\bibfield  {journal} {\bibinfo  {journal} {Nano
  letters}\ }\textbf {\bibinfo {volume} {16}},\ \bibinfo {pages} {825}
  (\bibinfo {year} {2016})}\BibitemShut {NoStop}%
\bibitem [{\citenamefont {Kang}\ \emph {et~al.}(2018)\citenamefont {Kang},
  \citenamefont {Fan}, \citenamefont {Zhi}, \citenamefont {Pan}, \citenamefont
  {Li}, \citenamefont {Wang}, \citenamefont {Guo}, \citenamefont {Zhao},\ and\
  \citenamefont {Xu}}]{kang2018two}%
  \BibitemOpen
  \bibfield  {author} {\bibinfo {author} {\bibfnamefont {N.}~\bibnamefont
  {Kang}}, \bibinfo {author} {\bibfnamefont {D.}~\bibnamefont {Fan}}, \bibinfo
  {author} {\bibfnamefont {J.}~\bibnamefont {Zhi}}, \bibinfo {author}
  {\bibfnamefont {D.}~\bibnamefont {Pan}}, \bibinfo {author} {\bibfnamefont
  {S.}~\bibnamefont {Li}}, \bibinfo {author} {\bibfnamefont {C.}~\bibnamefont
  {Wang}}, \bibinfo {author} {\bibfnamefont {J.}~\bibnamefont {Guo}}, \bibinfo
  {author} {\bibfnamefont {J.}~\bibnamefont {Zhao}},\ and\ \bibinfo {author}
  {\bibfnamefont {H.}~\bibnamefont {Xu}},\ }\bibfield  {title} {\bibinfo
  {title} {Two-dimensional quantum transport in free-standing {I}n{S}b
  nanosheets},\ }\href@noop {} {\bibfield  {journal} {\bibinfo  {journal} {Nano
  letters}\ }\textbf {\bibinfo {volume} {19}},\ \bibinfo {pages} {561}
  (\bibinfo {year} {2018})}\BibitemShut {NoStop}%
\bibitem [{\citenamefont {Xue}\ \emph {et~al.}(2019)\citenamefont {Xue},
  \citenamefont {Chen}, \citenamefont {Pan}, \citenamefont {Wang},
  \citenamefont {Zhao}, \citenamefont {Huang},\ and\ \citenamefont
  {Xu}}]{xue2019gate}%
  \BibitemOpen
  \bibfield  {author} {\bibinfo {author} {\bibfnamefont {J.}~\bibnamefont
  {Xue}}, \bibinfo {author} {\bibfnamefont {Y.}~\bibnamefont {Chen}}, \bibinfo
  {author} {\bibfnamefont {D.}~\bibnamefont {Pan}}, \bibinfo {author}
  {\bibfnamefont {J.-Y.}\ \bibnamefont {Wang}}, \bibinfo {author}
  {\bibfnamefont {J.}~\bibnamefont {Zhao}}, \bibinfo {author} {\bibfnamefont
  {S.}~\bibnamefont {Huang}},\ and\ \bibinfo {author} {\bibfnamefont
  {H.}~\bibnamefont {Xu}},\ }\bibfield  {title} {\bibinfo {title} {Gate defined
  quantum dot realized in a single crystalline {I}n{S}b nanosheet},\
  }\href@noop {} {\bibfield  {journal} {\bibinfo  {journal} {Applied Physics
  Letters}\ }\textbf {\bibinfo {volume} {114}},\ \bibinfo {pages} {023108}
  (\bibinfo {year} {2019})}\BibitemShut {NoStop}%
\bibitem [{\citenamefont {Lei}\ \emph {et~al.}(2020)\citenamefont {Lei},
  \citenamefont {Lehner}, \citenamefont {Rubi}, \citenamefont {Cheah},
  \citenamefont {Karalic}, \citenamefont {Mittag}, \citenamefont {Alt},
  \citenamefont {Scharnetzky}, \citenamefont {M\"arki}, \citenamefont
  {Zeitler}, \citenamefont {Wegscheider}, \citenamefont {Ihn},\ and\
  \citenamefont {Ensslin}}]{Lei2020}%
  \BibitemOpen
  \bibfield  {author} {\bibinfo {author} {\bibfnamefont {Z.}~\bibnamefont
  {Lei}}, \bibinfo {author} {\bibfnamefont {C.~A.}\ \bibnamefont {Lehner}},
  \bibinfo {author} {\bibfnamefont {K.}~\bibnamefont {Rubi}}, \bibinfo {author}
  {\bibfnamefont {E.}~\bibnamefont {Cheah}}, \bibinfo {author} {\bibfnamefont
  {M.}~\bibnamefont {Karalic}}, \bibinfo {author} {\bibfnamefont
  {C.}~\bibnamefont {Mittag}}, \bibinfo {author} {\bibfnamefont
  {L.}~\bibnamefont {Alt}}, \bibinfo {author} {\bibfnamefont {J.}~\bibnamefont
  {Scharnetzky}}, \bibinfo {author} {\bibfnamefont {P.}~\bibnamefont
  {M\"arki}}, \bibinfo {author} {\bibfnamefont {U.}~\bibnamefont {Zeitler}},
  \bibinfo {author} {\bibfnamefont {W.}~\bibnamefont {Wegscheider}}, \bibinfo
  {author} {\bibfnamefont {T.}~\bibnamefont {Ihn}},\ and\ \bibinfo {author}
  {\bibfnamefont {K.}~\bibnamefont {Ensslin}},\ }\bibfield  {title} {\bibinfo
  {title} {Electronic g factor and magnetotransport in {I}n{S}b quantum
  wells},\ }\href {https://doi.org/10.1103/PhysRevResearch.2.033213} {\bibfield
   {journal} {\bibinfo  {journal} {Physical Review Research}\ }\textbf
  {\bibinfo {volume} {2}},\ \bibinfo {pages} {033213} (\bibinfo {year}
  {2020})}\BibitemShut {NoStop}%
\bibitem [{\citenamefont {Lei}\ \emph {et~al.}(2021)\citenamefont {Lei},
  \citenamefont {Lehner}, \citenamefont {Cheah}, \citenamefont {Mittag},
  \citenamefont {Karalic}, \citenamefont {Wegscheider}, \citenamefont
  {Ensslin},\ and\ \citenamefont {Ihn}}]{Lei2021}%
  \BibitemOpen
  \bibfield  {author} {\bibinfo {author} {\bibfnamefont {Z.}~\bibnamefont
  {Lei}}, \bibinfo {author} {\bibfnamefont {C.~A.}\ \bibnamefont {Lehner}},
  \bibinfo {author} {\bibfnamefont {E.}~\bibnamefont {Cheah}}, \bibinfo
  {author} {\bibfnamefont {C.}~\bibnamefont {Mittag}}, \bibinfo {author}
  {\bibfnamefont {M.}~\bibnamefont {Karalic}}, \bibinfo {author} {\bibfnamefont
  {W.}~\bibnamefont {Wegscheider}}, \bibinfo {author} {\bibfnamefont
  {K.}~\bibnamefont {Ensslin}},\ and\ \bibinfo {author} {\bibfnamefont
  {T.}~\bibnamefont {Ihn}},\ }\bibfield  {title} {\bibinfo {title}
  {Gate-defined quantum point contact in an {I}n{S}b two-dimensional electron
  gas},\ }\href {https://doi.org/10.1103/PhysRevResearch.3.023042} {\bibfield
  {journal} {\bibinfo  {journal} {Physical Review Research}\ }\textbf {\bibinfo
  {volume} {3}},\ \bibinfo {pages} {023042} (\bibinfo {year}
  {2021})}\BibitemShut {NoStop}%
\bibitem [{\citenamefont {Lei}\ \emph {et~al.}(2022)\citenamefont {Lei},
  \citenamefont {Cheah}, \citenamefont {Rubi}, \citenamefont {Bal},
  \citenamefont {Adam}, \citenamefont {Schott}, \citenamefont {Zeitler},
  \citenamefont {Wegscheider}, \citenamefont {Ihn},\ and\ \citenamefont
  {Ensslin}}]{Lei2022}%
  \BibitemOpen
  \bibfield  {author} {\bibinfo {author} {\bibfnamefont {Z.}~\bibnamefont
  {Lei}}, \bibinfo {author} {\bibfnamefont {E.}~\bibnamefont {Cheah}}, \bibinfo
  {author} {\bibfnamefont {K.}~\bibnamefont {Rubi}}, \bibinfo {author}
  {\bibfnamefont {M.~E.}\ \bibnamefont {Bal}}, \bibinfo {author} {\bibfnamefont
  {C.}~\bibnamefont {Adam}}, \bibinfo {author} {\bibfnamefont {R.}~\bibnamefont
  {Schott}}, \bibinfo {author} {\bibfnamefont {U.}~\bibnamefont {Zeitler}},
  \bibinfo {author} {\bibfnamefont {W.}~\bibnamefont {Wegscheider}}, \bibinfo
  {author} {\bibfnamefont {T.}~\bibnamefont {Ihn}},\ and\ \bibinfo {author}
  {\bibfnamefont {K.}~\bibnamefont {Ensslin}},\ }\bibfield  {title} {\bibinfo
  {title} {High-quality two-dimensional electron gas in undoped {I}n{S}b
  quantum wells},\ }\href {https://doi.org/10.1103/PhysRevResearch.4.013039}
  {\bibfield  {journal} {\bibinfo  {journal} {Physical Review Research}\
  }\textbf {\bibinfo {volume} {4}},\ \bibinfo {pages} {013039} (\bibinfo {year}
  {2022})}\BibitemShut {NoStop}%
\bibitem [{\citenamefont {Eisenstein}\ \emph {et~al.}(1984)\citenamefont
  {Eisenstein}, \citenamefont {St\"ormer}, \citenamefont {Narayanamurti},
  \citenamefont {Gossard},\ and\ \citenamefont {Wiegmann}}]{PEisenstein1984}%
  \BibitemOpen
  \bibfield  {author} {\bibinfo {author} {\bibfnamefont {J.~P.}\ \bibnamefont
  {Eisenstein}}, \bibinfo {author} {\bibfnamefont {H.~L.}\ \bibnamefont
  {St\"ormer}}, \bibinfo {author} {\bibfnamefont {V.}~\bibnamefont
  {Narayanamurti}}, \bibinfo {author} {\bibfnamefont {A.~C.}\ \bibnamefont
  {Gossard}},\ and\ \bibinfo {author} {\bibfnamefont {W.}~\bibnamefont
  {Wiegmann}},\ }\bibfield  {title} {\bibinfo {title} {Effect of inversion
  symmetry on the band structure of semiconductor heterostructures},\ }\href
  {https://doi.org/10.1103/PhysRevLett.53.2579} {\bibfield  {journal} {\bibinfo
   {journal} {Physical Review Letters}\ }\textbf {\bibinfo {volume} {53}},\
  \bibinfo {pages} {2579} (\bibinfo {year} {1984})}\BibitemShut {NoStop}%
\bibitem [{\citenamefont {Marcellina}\ \emph {et~al.}(2017)\citenamefont
  {Marcellina}, \citenamefont {Hamilton}, \citenamefont {Winkler},\ and\
  \citenamefont {Culcer}}]{Marcellina2017}%
  \BibitemOpen
  \bibfield  {author} {\bibinfo {author} {\bibfnamefont {E.}~\bibnamefont
  {Marcellina}}, \bibinfo {author} {\bibfnamefont {A.~R.}\ \bibnamefont
  {Hamilton}}, \bibinfo {author} {\bibfnamefont {R.}~\bibnamefont {Winkler}},\
  and\ \bibinfo {author} {\bibfnamefont {D.}~\bibnamefont {Culcer}},\
  }\bibfield  {title} {\bibinfo {title} {Spin-orbit interactions in
  inversion-asymmetric two-dimensional hole systems: A variational analysis},\
  }\href {https://doi.org/10.1103/PhysRevB.95.075305} {\bibfield  {journal}
  {\bibinfo  {journal} {Physical Review B}\ }\textbf {\bibinfo {volume} {95}},\
  \bibinfo {pages} {075305} (\bibinfo {year} {2017})}\BibitemShut {NoStop}%
\bibitem [{\citenamefont {Pryor}\ and\ \citenamefont
  {Pistol}(2005)}]{Pryor2005}%
  \BibitemOpen
  \bibfield  {author} {\bibinfo {author} {\bibfnamefont {C.~E.}\ \bibnamefont
  {Pryor}}\ and\ \bibinfo {author} {\bibfnamefont {M.-E.}\ \bibnamefont
  {Pistol}},\ }\bibfield  {title} {\bibinfo {title} {Band-edge diagrams for
  strained {III}-{V} semiconductor quantum wells, wires, and dots},\ }\href
  {https://doi.org/10.1103/PhysRevB.72.205311} {\bibfield  {journal} {\bibinfo
  {journal} {Physical Review B}\ }\textbf {\bibinfo {volume} {72}},\ \bibinfo
  {pages} {205311} (\bibinfo {year} {2005})}\BibitemShut {NoStop}%
\bibitem [{\citenamefont {Gaspe}\ \emph {et~al.}(2011)\citenamefont {Gaspe},
  \citenamefont {Edirisooriya}, \citenamefont {Mishima}, \citenamefont
  {Jayathilaka}, \citenamefont {Doezema}, \citenamefont {Murphy}, \citenamefont
  {Santos}, \citenamefont {Tung},\ and\ \citenamefont {Wang}}]{Gaspe2011}%
  \BibitemOpen
  \bibfield  {author} {\bibinfo {author} {\bibfnamefont {C.~K.}\ \bibnamefont
  {Gaspe}}, \bibinfo {author} {\bibfnamefont {M.}~\bibnamefont {Edirisooriya}},
  \bibinfo {author} {\bibfnamefont {T.~D.}\ \bibnamefont {Mishima}}, \bibinfo
  {author} {\bibfnamefont {P.~A. R.~D.}\ \bibnamefont {Jayathilaka}}, \bibinfo
  {author} {\bibfnamefont {R.~E.}\ \bibnamefont {Doezema}}, \bibinfo {author}
  {\bibfnamefont {S.~Q.}\ \bibnamefont {Murphy}}, \bibinfo {author}
  {\bibfnamefont {M.~B.}\ \bibnamefont {Santos}}, \bibinfo {author}
  {\bibfnamefont {L.~C.}\ \bibnamefont {Tung}},\ and\ \bibinfo {author}
  {\bibfnamefont {Y.-J.}\ \bibnamefont {Wang}},\ }\bibfield  {title} {\bibinfo
  {title} {Effect of strain and confinement on the effective mass of holes in
  {I}n{S}b quantum wells},\ }\href {https://doi.org/10.1116/1.3553457}
  {\bibfield  {journal} {\bibinfo  {journal} {Journal of Vacuum Science \&
  Technology B, Nanotechnology and Microelectronics: Materials, Processing,
  Measurement, and Phenomena}\ }\textbf {\bibinfo {volume} {29}},\ \bibinfo
  {pages} {03C110} (\bibinfo {year} {2011})}\BibitemShut {NoStop}%
\bibitem [{\citenamefont {Nilsson}\ \emph {et~al.}(2011)\citenamefont
  {Nilsson}, \citenamefont {Deng}, \citenamefont {Caroff}, \citenamefont
  {Thelander}, \citenamefont {Samuelson}, \citenamefont {Wernersson},\ and\
  \citenamefont {Xu}}]{Nilsson2011}%
  \BibitemOpen
  \bibfield  {author} {\bibinfo {author} {\bibfnamefont {H.~A.}\ \bibnamefont
  {Nilsson}}, \bibinfo {author} {\bibfnamefont {M.~T.}\ \bibnamefont {Deng}},
  \bibinfo {author} {\bibfnamefont {P.}~\bibnamefont {Caroff}}, \bibinfo
  {author} {\bibfnamefont {C.}~\bibnamefont {Thelander}}, \bibinfo {author}
  {\bibfnamefont {L.}~\bibnamefont {Samuelson}}, \bibinfo {author}
  {\bibfnamefont {L.-E.}\ \bibnamefont {Wernersson}},\ and\ \bibinfo {author}
  {\bibfnamefont {H.~Q.}\ \bibnamefont {Xu}},\ }\bibfield  {title} {\bibinfo
  {title} {{I}n{S}b nanowire field-effect transistors and quantum-dot
  devices},\ }\href {https://doi.org/10.1109/JSTQE.2010.2090135} {\bibfield
  {journal} {\bibinfo  {journal} {IEEE Journal of Selected Topics in Quantum
  Electronics}\ }\textbf {\bibinfo {volume} {17}},\ \bibinfo {pages} {907}
  (\bibinfo {year} {2011})}\BibitemShut {NoStop}%
\bibitem [{\citenamefont {Pribiag}\ \emph {et~al.}(2013)\citenamefont
  {Pribiag}, \citenamefont {Nadj-Perge}, \citenamefont {Frolov}, \citenamefont
  {Berg}, \citenamefont {Weperen}, \citenamefont {Plissard}, \citenamefont
  {Bakkers},\ and\ \citenamefont {Kouwenhoven}}]{Pribiag2013}%
  \BibitemOpen
  \bibfield  {author} {\bibinfo {author} {\bibfnamefont {V.~S.}\ \bibnamefont
  {Pribiag}}, \bibinfo {author} {\bibfnamefont {S.}~\bibnamefont {Nadj-Perge}},
  \bibinfo {author} {\bibfnamefont {S.~M.}\ \bibnamefont {Frolov}}, \bibinfo
  {author} {\bibfnamefont {J.~W. V.~D.}\ \bibnamefont {Berg}}, \bibinfo
  {author} {\bibfnamefont {I.~V.}\ \bibnamefont {Weperen}}, \bibinfo {author}
  {\bibfnamefont {S.~R.}\ \bibnamefont {Plissard}}, \bibinfo {author}
  {\bibfnamefont {E.~P.}\ \bibnamefont {Bakkers}},\ and\ \bibinfo {author}
  {\bibfnamefont {L.~P.}\ \bibnamefont {Kouwenhoven}},\ }\bibfield  {title}
  {\bibinfo {title} {Electrical control of single hole spins in nanowire
  quantum dots},\ }\href {https://doi.org/10.1038/nnano.2013.5} {\bibfield
  {journal} {\bibinfo  {journal} {Nature Nanotechnology}\ }\textbf {\bibinfo
  {volume} {8}},\ \bibinfo {pages} {170} (\bibinfo {year} {2013})}\BibitemShut
  {NoStop}%
\bibitem [{\citenamefont {Radosavljevic}\ \emph {et~al.}(2008)\citenamefont
  {Radosavljevic}, \citenamefont {Ashley}, \citenamefont {Andreev},
  \citenamefont {Coomber}, \citenamefont {Dewey}, \citenamefont {Emeny},
  \citenamefont {Fearn}, \citenamefont {Hayes}, \citenamefont {Hilton},
  \citenamefont {Hudait} \emph {et~al.}}]{radosavljevic2008high}%
  \BibitemOpen
  \bibfield  {author} {\bibinfo {author} {\bibfnamefont {M.}~\bibnamefont
  {Radosavljevic}}, \bibinfo {author} {\bibfnamefont {T.}~\bibnamefont
  {Ashley}}, \bibinfo {author} {\bibfnamefont {A.}~\bibnamefont {Andreev}},
  \bibinfo {author} {\bibfnamefont {S.}~\bibnamefont {Coomber}}, \bibinfo
  {author} {\bibfnamefont {G.}~\bibnamefont {Dewey}}, \bibinfo {author}
  {\bibfnamefont {M.}~\bibnamefont {Emeny}}, \bibinfo {author} {\bibfnamefont
  {M.}~\bibnamefont {Fearn}}, \bibinfo {author} {\bibfnamefont
  {D.}~\bibnamefont {Hayes}}, \bibinfo {author} {\bibfnamefont
  {K.}~\bibnamefont {Hilton}}, \bibinfo {author} {\bibfnamefont
  {M.}~\bibnamefont {Hudait}}, \emph {et~al.},\ }\bibfield  {title} {\bibinfo
  {title} {High-performance 40nm gate length {I}n{S}b p-channel compressively
  strained quantum well field effect transistors for low-power ({VCC}= 0.5 {V})
  logic applications},\ }in\ \href@noop {} {\emph {\bibinfo {booktitle} {2008
  IEEE International Electron Devices Meeting}}}\ (\bibinfo {organization}
  {IEEE},\ \bibinfo {year} {2008})\ pp.\ \bibinfo {pages} {1--4}\BibitemShut
  {NoStop}%
\bibitem [{\citenamefont {Winkler}(2003)}]{Winkler2003}%
  \BibitemOpen
  \bibfield  {author} {\bibinfo {author} {\bibfnamefont {R.}~\bibnamefont
  {Winkler}},\ }\href@noop {} {\emph {\bibinfo {title} {Spin-orbit coupling
  effects in twoi-dimensional electron and hole systems}}}\ (\bibinfo
  {publisher} {Springer, Berlin},\ \bibinfo {year} {2003})\BibitemShut
  {NoStop}%
\bibitem [{\citenamefont {Ihn}(2010)}]{Ihn}%
  \BibitemOpen
  \bibfield  {author} {\bibinfo {author} {\bibfnamefont {T.}~\bibnamefont
  {Ihn}},\ }\href@noop {} {\emph {\bibinfo {title} {Semiconductor
  Nanostructures: Quantum States and Electronic Transport}}}\ (\bibinfo
  {publisher} {Oxford University Press},\ \bibinfo {year} {2010})\BibitemShut
  {NoStop}%
\bibitem [{\citenamefont {Umansky}\ \emph {et~al.}(1997)\citenamefont
  {Umansky}, \citenamefont {De-Picciotto},\ and\ \citenamefont
  {Heiblum}}]{umansky1997extremely}%
  \BibitemOpen
  \bibfield  {author} {\bibinfo {author} {\bibfnamefont {V.}~\bibnamefont
  {Umansky}}, \bibinfo {author} {\bibfnamefont {R.}~\bibnamefont
  {De-Picciotto}},\ and\ \bibinfo {author} {\bibfnamefont {M.}~\bibnamefont
  {Heiblum}},\ }\bibfield  {title} {\bibinfo {title} {Extremely high-mobility
  two dimensional electron gas: Evaluation of scattering mechanisms},\
  }\href@noop {} {\bibfield  {journal} {\bibinfo  {journal} {Applied Physics
  Letters}\ }\textbf {\bibinfo {volume} {71}},\ \bibinfo {pages} {683}
  (\bibinfo {year} {1997})}\BibitemShut {NoStop}%
\bibitem [{\citenamefont {Davies}(1993)}]{Davies1997}%
  \BibitemOpen
  \bibfield  {author} {\bibinfo {author} {\bibfnamefont {J.~H.}\ \bibnamefont
  {Davies}},\ }\href@noop {} {\emph {\bibinfo {title} {The physics of
  low-dimensional semiconductors, an introduction}}}\ (\bibinfo  {publisher}
  {Cambridge University press},\ \bibinfo {year} {1993})\BibitemShut {NoStop}%
\bibitem [{\citenamefont {Chung}\ \emph {et~al.}(2022)\citenamefont {Chung},
  \citenamefont {Wang}, \citenamefont {Singh}, \citenamefont {Gupta},
  \citenamefont {Baldwin}, \citenamefont {West}, \citenamefont {Shayegan},
  \citenamefont {Pfeiffer},\ and\ \citenamefont {Winkler}}]{Chung2022}%
  \BibitemOpen
  \bibfield  {author} {\bibinfo {author} {\bibfnamefont {Y.~J.}\ \bibnamefont
  {Chung}}, \bibinfo {author} {\bibfnamefont {C.}~\bibnamefont {Wang}},
  \bibinfo {author} {\bibfnamefont {S.~K.}\ \bibnamefont {Singh}}, \bibinfo
  {author} {\bibfnamefont {A.}~\bibnamefont {Gupta}}, \bibinfo {author}
  {\bibfnamefont {K.~W.}\ \bibnamefont {Baldwin}}, \bibinfo {author}
  {\bibfnamefont {K.~W.}\ \bibnamefont {West}}, \bibinfo {author}
  {\bibfnamefont {M.}~\bibnamefont {Shayegan}}, \bibinfo {author}
  {\bibfnamefont {L.~N.}\ \bibnamefont {Pfeiffer}},\ and\ \bibinfo {author}
  {\bibfnamefont {R.}~\bibnamefont {Winkler}},\ }\bibfield  {title} {\bibinfo
  {title} {Record-quality {G}a{A}s two-dimensional hole systems},\ }\href
  {https://doi.org/10.1103/PhysRevMaterials.6.034005} {\bibfield  {journal}
  {\bibinfo  {journal} {Physical Review Materials}\ }\textbf {\bibinfo {volume}
  {6}},\ \bibinfo {pages} {034005} (\bibinfo {year} {2022})}\BibitemShut
  {NoStop}%
\bibitem [{\citenamefont {Yi}\ \emph {et~al.}(2015)\citenamefont {Yi},
  \citenamefont {Kiselev}, \citenamefont {Thorp}, \citenamefont {Noah},
  \citenamefont {Nguyen}, \citenamefont {Bui}, \citenamefont {Rajavel},
  \citenamefont {Hussain}, \citenamefont {Gyure}, \citenamefont {Kratz} \emph
  {et~al.}}]{yi2015gate}%
  \BibitemOpen
  \bibfield  {author} {\bibinfo {author} {\bibfnamefont {W.}~\bibnamefont
  {Yi}}, \bibinfo {author} {\bibfnamefont {A.~A.}\ \bibnamefont {Kiselev}},
  \bibinfo {author} {\bibfnamefont {J.}~\bibnamefont {Thorp}}, \bibinfo
  {author} {\bibfnamefont {R.}~\bibnamefont {Noah}}, \bibinfo {author}
  {\bibfnamefont {B.-M.}\ \bibnamefont {Nguyen}}, \bibinfo {author}
  {\bibfnamefont {S.}~\bibnamefont {Bui}}, \bibinfo {author} {\bibfnamefont
  {R.~D.}\ \bibnamefont {Rajavel}}, \bibinfo {author} {\bibfnamefont
  {T.}~\bibnamefont {Hussain}}, \bibinfo {author} {\bibfnamefont {M.~F.}\
  \bibnamefont {Gyure}}, \bibinfo {author} {\bibfnamefont {P.}~\bibnamefont
  {Kratz}}, \emph {et~al.},\ }\bibfield  {title} {\bibinfo {title}
  {Gate-tunable high mobility remote-doped {I}n{S}b/in1- xalxsb quantum well
  heterostructures},\ }\href@noop {} {\bibfield  {journal} {\bibinfo  {journal}
  {Applied Physics Letters}\ }\textbf {\bibinfo {volume} {106}},\ \bibinfo
  {pages} {142103} (\bibinfo {year} {2015})}\BibitemShut {NoStop}%
\bibitem [{\citenamefont {Tschirky}\ \emph {et~al.}(2017)\citenamefont
  {Tschirky}, \citenamefont {Mueller}, \citenamefont {Lehner}, \citenamefont
  {Faelt}, \citenamefont {Ihn}, \citenamefont {Ensslin},\ and\ \citenamefont
  {Wegscheider}}]{Tschirky2017}%
  \BibitemOpen
  \bibfield  {author} {\bibinfo {author} {\bibfnamefont {T.}~\bibnamefont
  {Tschirky}}, \bibinfo {author} {\bibfnamefont {S.}~\bibnamefont {Mueller}},
  \bibinfo {author} {\bibfnamefont {C.~A.}\ \bibnamefont {Lehner}}, \bibinfo
  {author} {\bibfnamefont {S.}~\bibnamefont {Faelt}}, \bibinfo {author}
  {\bibfnamefont {T.}~\bibnamefont {Ihn}}, \bibinfo {author} {\bibfnamefont
  {K.}~\bibnamefont {Ensslin}},\ and\ \bibinfo {author} {\bibfnamefont
  {W.}~\bibnamefont {Wegscheider}},\ }\bibfield  {title} {\bibinfo {title}
  {Scattering mechanisms of highest-mobility
  {I}n{A}s$/${A}l$_{x}${G}a$_{1-x}${S}b quantum wells},\ }\href
  {https://doi.org/10.1103/PhysRevB.95.115304} {\bibfield  {journal} {\bibinfo
  {journal} {Physical Review B}\ }\textbf {\bibinfo {volume} {95}},\ \bibinfo
  {pages} {115304} (\bibinfo {year} {2017})}\BibitemShut {NoStop}%
\bibitem [{\citenamefont {Ahn}\ and\ \citenamefont
  {Das~Sarma}(2022)}]{Ahn2022}%
  \BibitemOpen
  \bibfield  {author} {\bibinfo {author} {\bibfnamefont {S.}~\bibnamefont
  {Ahn}}\ and\ \bibinfo {author} {\bibfnamefont {S.}~\bibnamefont
  {Das~Sarma}},\ }\bibfield  {title} {\bibinfo {title} {Density-dependent
  two-dimensional optimal mobility in ultra-high-quality semiconductor quantum
  wells},\ }\href {https://doi.org/10.1103/PhysRevMaterials.6.014603}
  {\bibfield  {journal} {\bibinfo  {journal} {Physical Review Materials}\
  }\textbf {\bibinfo {volume} {6}},\ \bibinfo {pages} {014603} (\bibinfo {year}
  {2022})}\BibitemShut {NoStop}%
\bibitem [{\citenamefont {Hikami}\ \emph {et~al.}(1980)\citenamefont {Hikami},
  \citenamefont {Larkin},\ and\ \citenamefont {Nagaoka}}]{Hikami1980}%
  \BibitemOpen
  \bibfield  {author} {\bibinfo {author} {\bibfnamefont {S.}~\bibnamefont
  {Hikami}}, \bibinfo {author} {\bibfnamefont {A.~I.}\ \bibnamefont {Larkin}},\
  and\ \bibinfo {author} {\bibfnamefont {Y.}~\bibnamefont {Nagaoka}},\
  }\bibfield  {title} {\bibinfo {title} {{Spin-Orbit Interaction and
  Magnetoresistance in the Two Dimensional Random System}},\ }\href
  {https://doi.org/10.1143/PTP.63.707} {\bibfield  {journal} {\bibinfo
  {journal} {Progress of Theoretical Physics}\ }\textbf {\bibinfo {volume}
  {63}},\ \bibinfo {pages} {707} (\bibinfo {year} {1980})},\ \Eprint
  {https://arxiv.org/abs/https://academic.oup.com/ptp/article-pdf/63/2/707/5336056/63-2-707.pdf}
  {https://academic.oup.com/ptp/article-pdf/63/2/707/5336056/63-2-707.pdf}
  \BibitemShut {NoStop}%
\bibitem [{\citenamefont {V.}\ \emph {et~al.}(1994)\citenamefont {V.},
  \citenamefont {B.},\ and\ \citenamefont {E.}}]{Iordanskii1994}%
  \BibitemOpen
  \bibfield  {author} {\bibinfo {author} {\bibfnamefont {I.~S.}\ \bibnamefont
  {V.}}, \bibinfo {author} {\bibfnamefont {L.-G.~Y.}\ \bibnamefont {B.}},\ and\
  \bibinfo {author} {\bibfnamefont {P.~G.}\ \bibnamefont {E.}},\ }\bibfield
  {title} {\bibinfo {title} {{Weak localization in quantum wells with
  spin-orbit interaction}},\ }\href
  {http://jetpletters.ru/ps/1323/article_20010.shtml} {\bibfield  {journal}
  {\bibinfo  {journal} {Pis’ma Zh. Eksp. Teor. Fiz.}\ }\textbf {\bibinfo
  {volume} {60}} (\bibinfo {year} {1994})}\BibitemShut {NoStop}%
\bibitem [{\citenamefont {Chen}\ \emph {et~al.}(2021)\citenamefont {Chen},
  \citenamefont {Huang}, \citenamefont {Pan}, \citenamefont {Xue},
  \citenamefont {Zhang}, \citenamefont {Zhao},\ and\ \citenamefont
  {Xu}}]{chen2021strong}%
  \BibitemOpen
  \bibfield  {author} {\bibinfo {author} {\bibfnamefont {Y.}~\bibnamefont
  {Chen}}, \bibinfo {author} {\bibfnamefont {S.}~\bibnamefont {Huang}},
  \bibinfo {author} {\bibfnamefont {D.}~\bibnamefont {Pan}}, \bibinfo {author}
  {\bibfnamefont {J.}~\bibnamefont {Xue}}, \bibinfo {author} {\bibfnamefont
  {L.}~\bibnamefont {Zhang}}, \bibinfo {author} {\bibfnamefont
  {J.}~\bibnamefont {Zhao}},\ and\ \bibinfo {author} {\bibfnamefont
  {H.}~\bibnamefont {Xu}},\ }\bibfield  {title} {\bibinfo {title} {Strong and
  tunable spin--orbit interaction in a single crystalline {I}n{S}b nanosheet},\
  }\href@noop {} {\bibfield  {journal} {\bibinfo  {journal} {npj 2D Materials
  and Applications}\ }\textbf {\bibinfo {volume} {5}},\ \bibinfo {pages} {1}
  (\bibinfo {year} {2021})}\BibitemShut {NoStop}%
\bibitem [{\citenamefont {Senz}\ \emph {et~al.}(2000)\citenamefont {Senz},
  \citenamefont {Heinzel}, \citenamefont {Ihn}, \citenamefont {Ensslin},
  \citenamefont {Dehlinger}, \citenamefont {Gr\"utzmacher},\ and\ \citenamefont
  {Gennser}}]{Senz2000}%
  \BibitemOpen
  \bibfield  {author} {\bibinfo {author} {\bibfnamefont {V.}~\bibnamefont
  {Senz}}, \bibinfo {author} {\bibfnamefont {T.}~\bibnamefont {Heinzel}},
  \bibinfo {author} {\bibfnamefont {T.}~\bibnamefont {Ihn}}, \bibinfo {author}
  {\bibfnamefont {K.}~\bibnamefont {Ensslin}}, \bibinfo {author} {\bibfnamefont
  {G.}~\bibnamefont {Dehlinger}}, \bibinfo {author} {\bibfnamefont
  {D.}~\bibnamefont {Gr\"utzmacher}},\ and\ \bibinfo {author} {\bibfnamefont
  {U.}~\bibnamefont {Gennser}},\ }\bibfield  {title} {\bibinfo {title}
  {Coexistence of weak localization and a metallic phase in {S}i/{S}i{G}e
  quantum wells},\ }\href {https://doi.org/10.1103/PhysRevB.61.R5082}
  {\bibfield  {journal} {\bibinfo  {journal} {Physical Review B}\ }\textbf
  {\bibinfo {volume} {61}},\ \bibinfo {pages} {R5082} (\bibinfo {year}
  {2000})}\BibitemShut {NoStop}%
\bibitem [{\citenamefont {Grbi\ifmmode~\acute{c}\else \'{c}\fi{}}\ \emph
  {et~al.}(2008)\citenamefont {Grbi\ifmmode~\acute{c}\else \'{c}\fi{}},
  \citenamefont {Leturcq}, \citenamefont {Ihn}, \citenamefont {Ensslin},
  \citenamefont {Reuter},\ and\ \citenamefont {Wieck}}]{Grbic2008}%
  \BibitemOpen
  \bibfield  {author} {\bibinfo {author} {\bibfnamefont {B.}~\bibnamefont
  {Grbi\ifmmode~\acute{c}\else \'{c}\fi{}}}, \bibinfo {author} {\bibfnamefont
  {R.}~\bibnamefont {Leturcq}}, \bibinfo {author} {\bibfnamefont
  {T.}~\bibnamefont {Ihn}}, \bibinfo {author} {\bibfnamefont {K.}~\bibnamefont
  {Ensslin}}, \bibinfo {author} {\bibfnamefont {D.}~\bibnamefont {Reuter}},\
  and\ \bibinfo {author} {\bibfnamefont {A.~D.}\ \bibnamefont {Wieck}},\
  }\bibfield  {title} {\bibinfo {title} {Strong spin-orbit interactions and
  weak antilocalization in carbon-doped $p$-type
  {G}a{A}s/{A}l$_{x}${G}a$_{1-x}${A}s heterostructures},\ }\href
  {https://doi.org/10.1103/PhysRevB.77.125312} {\bibfield  {journal} {\bibinfo
  {journal} {Physical Review B}\ }\textbf {\bibinfo {volume} {77}},\ \bibinfo
  {pages} {125312} (\bibinfo {year} {2008})}\BibitemShut {NoStop}%
\bibitem [{\citenamefont {Metti}\ \emph {et~al.}(2022)\citenamefont {Metti},
  \citenamefont {Thomas}, \citenamefont {Xiao},\ and\ \citenamefont
  {Manfra}}]{metti2022}%
  \BibitemOpen
  \bibfield  {author} {\bibinfo {author} {\bibfnamefont {S.}~\bibnamefont
  {Metti}}, \bibinfo {author} {\bibfnamefont {C.}~\bibnamefont {Thomas}},
  \bibinfo {author} {\bibfnamefont {D.}~\bibnamefont {Xiao}},\ and\ \bibinfo
  {author} {\bibfnamefont {M.}~\bibnamefont {Manfra}},\ }\bibfield  {title}
  {\bibinfo {title} {Spin-orbit coupling and electron scattering in
  high-quality {I}n{S}b$_{1-x}${A}s$_{x}$ quantum wells},\ }\href@noop {}
  {\bibfield  {journal} {\bibinfo  {journal} {arXiv preprint arXiv:2207.02006}\
  } (\bibinfo {year} {2022})}\BibitemShut {NoStop}%
\bibitem [{\citenamefont {Brosig}\ \emph {et~al.}(2000)\citenamefont {Brosig},
  \citenamefont {Ensslin}, \citenamefont {Jansen}, \citenamefont {Nguyen},
  \citenamefont {Brar}, \citenamefont {Thomas},\ and\ \citenamefont
  {Kroemer}}]{Brossig2000}%
  \BibitemOpen
  \bibfield  {author} {\bibinfo {author} {\bibfnamefont {S.}~\bibnamefont
  {Brosig}}, \bibinfo {author} {\bibfnamefont {K.}~\bibnamefont {Ensslin}},
  \bibinfo {author} {\bibfnamefont {A.~G.}\ \bibnamefont {Jansen}}, \bibinfo
  {author} {\bibfnamefont {C.}~\bibnamefont {Nguyen}}, \bibinfo {author}
  {\bibfnamefont {B.}~\bibnamefont {Brar}}, \bibinfo {author} {\bibfnamefont
  {M.}~\bibnamefont {Thomas}},\ and\ \bibinfo {author} {\bibfnamefont
  {H.}~\bibnamefont {Kroemer}},\ }\bibfield  {title} {\bibinfo {title}
  {{I}n{A}s-alsb quantum wells in tilted magnetic fields},\ }\href
  {https://doi.org/10.1103/PhysRevB.61.13045} {\bibfield  {journal} {\bibinfo
  {journal} {Physical Review B}\ }\textbf {\bibinfo {volume} {61}},\ \bibinfo
  {pages} {13045} (\bibinfo {year} {2000})}\BibitemShut {NoStop}%
\bibitem [{\citenamefont {Schumacher}\ \emph {et~al.}(1998)\citenamefont
  {Schumacher}, \citenamefont {Nauen}, \citenamefont {Zeitler}, \citenamefont
  {Haug}, \citenamefont {Weitz}, \citenamefont {Jansen},\ and\ \citenamefont
  {Schäffler}}]{SCHUMACHER1998}%
  \BibitemOpen
  \bibfield  {author} {\bibinfo {author} {\bibfnamefont {H.}~\bibnamefont
  {Schumacher}}, \bibinfo {author} {\bibfnamefont {A.}~\bibnamefont {Nauen}},
  \bibinfo {author} {\bibfnamefont {U.}~\bibnamefont {Zeitler}}, \bibinfo
  {author} {\bibfnamefont {R.}~\bibnamefont {Haug}}, \bibinfo {author}
  {\bibfnamefont {P.}~\bibnamefont {Weitz}}, \bibinfo {author} {\bibfnamefont
  {A.}~\bibnamefont {Jansen}},\ and\ \bibinfo {author} {\bibfnamefont
  {F.}~\bibnamefont {Schäffler}},\ }\bibfield  {title} {\bibinfo {title}
  {Anomalous coincidences between valley split landau levels in a {S}i/{S}i{G}e
  heterostructure},\ }\href
  {https://doi.org/https://doi.org/10.1016/S0921-4526(98)00589-4} {\bibfield
  {journal} {\bibinfo  {journal} {Physica B: Condensed Matter}\ }\textbf
  {\bibinfo {volume} {256-258}},\ \bibinfo {pages} {260} (\bibinfo {year}
  {1998})}\BibitemShut {NoStop}%
\bibitem [{\citenamefont {Haug}\ \emph {et~al.}(1987)\citenamefont {Haug},
  \citenamefont {Gerhardts}, \citenamefont {Klitzing},\ and\ \citenamefont
  {Ploog}}]{Raug1987}%
  \BibitemOpen
  \bibfield  {author} {\bibinfo {author} {\bibfnamefont {R.~J.}\ \bibnamefont
  {Haug}}, \bibinfo {author} {\bibfnamefont {R.~R.}\ \bibnamefont {Gerhardts}},
  \bibinfo {author} {\bibfnamefont {K.~v.}\ \bibnamefont {Klitzing}},\ and\
  \bibinfo {author} {\bibfnamefont {K.}~\bibnamefont {Ploog}},\ }\bibfield
  {title} {\bibinfo {title} {Effect of repulsive and attractive scattering
  centers on the magnetotransport properties of a two-dimensional electron
  gas},\ }\href {https://doi.org/10.1103/PhysRevLett.59.1349} {\bibfield
  {journal} {\bibinfo  {journal} {Physical Review Letters}\ }\textbf {\bibinfo
  {volume} {59}},\ \bibinfo {pages} {1349} (\bibinfo {year}
  {1987})}\BibitemShut {NoStop}%
\bibitem [{\citenamefont {Shi}\ \emph {et~al.}(2015)\citenamefont {Shi},
  \citenamefont {Zudov}, \citenamefont {Morrison},\ and\ \citenamefont
  {Myronov}}]{Shi2015}%
  \BibitemOpen
  \bibfield  {author} {\bibinfo {author} {\bibfnamefont {Q.}~\bibnamefont
  {Shi}}, \bibinfo {author} {\bibfnamefont {M.~A.}\ \bibnamefont {Zudov}},
  \bibinfo {author} {\bibfnamefont {C.}~\bibnamefont {Morrison}},\ and\
  \bibinfo {author} {\bibfnamefont {M.}~\bibnamefont {Myronov}},\ }\bibfield
  {title} {\bibinfo {title} {Strong transport anisotropy in {G}e/{S}i{G}e
  quantum wells in tilted magnetic fields},\ }\href
  {https://doi.org/10.1103/PhysRevB.91.201301} {\bibfield  {journal} {\bibinfo
  {journal} {Physical Review B}\ }\textbf {\bibinfo {volume} {91}},\ \bibinfo
  {pages} {201301} (\bibinfo {year} {2015})}\BibitemShut {NoStop}%
\bibitem [{\citenamefont {Pisoni}\ \emph {et~al.}(2017)\citenamefont {Pisoni},
  \citenamefont {Lee}, \citenamefont {Overweg}, \citenamefont {Eich},
  \citenamefont {Simonet}, \citenamefont {Watanabe}, \citenamefont {Taniguchi},
  \citenamefont {Gorbachev}, \citenamefont {Ihn},\ and\ \citenamefont
  {Ensslin}}]{Pisoni2017}%
  \BibitemOpen
  \bibfield  {author} {\bibinfo {author} {\bibfnamefont {R.}~\bibnamefont
  {Pisoni}}, \bibinfo {author} {\bibfnamefont {Y.}~\bibnamefont {Lee}},
  \bibinfo {author} {\bibfnamefont {H.}~\bibnamefont {Overweg}}, \bibinfo
  {author} {\bibfnamefont {M.}~\bibnamefont {Eich}}, \bibinfo {author}
  {\bibfnamefont {P.}~\bibnamefont {Simonet}}, \bibinfo {author} {\bibfnamefont
  {K.}~\bibnamefont {Watanabe}}, \bibinfo {author} {\bibfnamefont
  {T.}~\bibnamefont {Taniguchi}}, \bibinfo {author} {\bibfnamefont
  {R.}~\bibnamefont {Gorbachev}}, \bibinfo {author} {\bibfnamefont
  {T.}~\bibnamefont {Ihn}},\ and\ \bibinfo {author} {\bibfnamefont
  {K.}~\bibnamefont {Ensslin}},\ }\bibfield  {title} {\bibinfo {title}
  {Gate-defined one-dimensional channel and broken symmetry states in
  {M}o{S}$_2$ van der {W}aals heterostructures},\ }\href
  {https://doi.org/10.1021/acs.nanolett.7b02186} {\bibfield  {journal}
  {\bibinfo  {journal} {Nano Letters}\ }\textbf {\bibinfo {volume} {17}},\
  \bibinfo {pages} {5008} (\bibinfo {year} {2017})},\ \bibinfo {note} {pMID:
  28686030},\ \Eprint
  {https://arxiv.org/abs/https://doi.org/10.1021/acs.nanolett.7b02186}
  {https://doi.org/10.1021/acs.nanolett.7b02186} \BibitemShut {NoStop}%
\end{thebibliography}%
\end{document}